\documentclass[pra,twocolumn,showpacs,preprintnumbers,superscriptaddress,amsmath,amssymb]{revtex4}
\usepackage{graphicx}
\usepackage{bm}
\usepackage{psfrag}
\usepackage{epsfig}
\usepackage{amsmath}
\usepackage{amssymb}
\usepackage{color}
\usepackage{subfigure}
\newcommand{\nn}{\nonumber}

\begin{document}

\title{$GHZ$ and $W$ entanglement witnesses for the noninteracting Fermi gas}

\author{Hessam Habibian}\thanks{E-mail: hessam.habibian@uab.es}
\affiliation{Christian Doppler Labor f\"{u}r
Oberfl\"{a}chenoptische Methoden, Johannes Kepler Universit\"{a}t,
Linz A-4040, Austria}
\affiliation{Grup d'\'{O}ptica, Departament
de F\'{i}sica, Universitat Aut\`{o}noma de Barcelona, Bellaterra
E-08193, Barcelona, Spain}

\author{John W. Clark}
\affiliation{Institut f\"ur Theoretische Physik, Johannes Kepler
Universit\"{a}t, Linz A-4040, Austria} \affiliation{Department of
Physics, CB 1105, Washington University, St. Louis, Missouri
63130, USA}

\author{Naeimeh Behbood}
\affiliation{Christian Doppler Labor f\"{u}r
Oberfl\"{a}chenoptische Methoden, Johannes Kepler Universit\"{a}t,
Linz A-4040, Austria}
\affiliation{ICFO-Institut de Ciencies
Fotoniques, Mediterranean Technology Park, Castelldefels E-08860,
Barcelona, Spain}

\author{Kurt Hingerl}
\affiliation{Christian Doppler Labor f\"{u}r
Oberfl\"{a}chenoptische Methoden, Johannes Kepler Universit\"{a}t,
Linz A-4040, Austria}
\date{\today}

\begin{abstract}

The existence and nature of tripartite entanglement of a
noninteracting Fermi gas (NIFG) is investigated. Three new classes
of parameterized entanglement witnesses (EWs) are introduced with
the aim of detecting genuine tripartite entanglement in the
three-body reduced density matrix and discriminating between the
presence of the two types of genuine tripartite entanglement,
$W\setminus B$ and $GHZ\setminus W$. By choosing appropriate EW
operators, the problem of finding $GHZ$ and $W$ EWs is reduced to
linear programming. Specifically, we devise new $W$ EWs based on a
spin-chain model with periodic boundary conditions, and we
construct a class of parametrized $GHZ$ EWs by linearly combining
projection operators corresponding to all the different
state-vector types arising for a three-fermion system. A third
class of EWs is provided by a $GHZ$ stabilizer operator capable of
distinguishing $W\setminus B$ from $GHZ\setminus B$ entanglement,
which is not possible with $W$ EWs. Implementing these classes of
EWs, it is found that all states containing genuine tripartite
entanglement are of $W$ type, and hence states containing $GHZ
\setminus W$ genuine tripartite entanglement do not arise. Some
genuine tripartite entangled states that have a positive partial
transpose (PPT) with respect to some bipartition are detected.
Finally, it is demonstrated that a NIFG does not exhibit ``pure''
$W\setminus B$ genuine tripartite entanglement: three-party
entanglement without any separable or biseparable admixture does
not occur.
\end{abstract}

\pacs{03.67.Mn, 03.65.Ud, 71.10.Ca}

\maketitle

\section{Introduction}

The existence of entangled states is a distinctive signature of
quantum mechanics of the most profound conceptual and practical
importance.  The phenomenon of entanglement promises to be the
source of many far-reaching technological advances of the 21st
century.  A fundamental and quantitative understanding of its nature
will be instrumental to its exploitation in quantum information
processing based on optical and condensed-matter systems.  Moreover,
such an understanding can bring new insights into the microscopic
physics, statistical mechanics, and phenomenology of strongly
interacting quantum many-body systems.

There is a growing body of work that seeks to establish the
entanglement properties of the states of quantum many-body systems
and the roles entanglement plays in the observed behavior of these
systems.  In particular, much progress has been made on the
entanglement properties of spin-lattice models, stimulated by the
pioneering studies of Osterloh \emph{et al.}\ \cite{Ost} and
Osborne and Nielsen \cite{ON}. The recent review of Amico \emph{et
al.}\ \cite{Am1} assesses the state of knowledge on bipartite and
multipartite entanglement for diverse many-body systems including
spin, fermion, and boson models.  The great majority of examples
studied involve spin systems and systems of particles made
distinguishable by localization.

There is an ongoing debate on the nature of entanglement in
systems of identical particles---just how does the
indistinguishability of the particles impact the quantification of
entanglement?  Attempts at clarification of the various issues
that arise when different Bose and Fermi systems with different
degrees of freedom are studied \cite{Am1, Schl, Eck1, Ghir1, Ved2,
zan, Ved1, Oh, Lunk, Dow, Vert1} has led to the examination of
various quantities deemed to measure or detect entanglement in the
presence of indistinguishability. Here we shall (i) focus on the
noninteracting Fermi gas (NIFG) as represented by the
three-fermion reduced density matrix of its ground state and (ii)
adopt the entanglement witness (EW) criterion
\cite{Lewenstein,Acin} for analysis of the entanglement content of
this tripartite state descriptor, which overcomes disadvantages of
some of the earlier work. We shall introduce new classes of
parameterized EW operators for indistinguishable fermions, to
enable detection of $GHZ\setminus W$ (i.e., the subset of $GHZ$
subtracted by the set of $W$) and $W\setminus B$ genuine
tripartite entanglement in the NIFG, if one or the other is
present.

This paper is organized as follows: In Sec.~II we review certain
definitions basic to the discussion of entanglement in systems of
distinguishable or indistinguishable particles, and some existing
formalism and results from other authors relevant to
characterization of the entanglement properties of the
noninteracting Fermi gas \cite{Schl,Eck1,Ved1,Oh,Lunk}.
Importantly, we display the general form obtained for the
three-particle reduced density matrix of the NIFG, which will be
the central quantity of our analysis.  In Sec.~III we consider the
classes of tripartite entanglement identified for mixed
three-qubit states by Ac\'in \emph{et al.}\ \cite{Acin} and
discuss the properties of entanglement witnesses (EWs) capable of
signaling the presence of these classes.  A general scheme for
constructing parametrized $GHZ$ and $W$ EWs via linear programming
(LP) is introduced as a special case of convex optimization.
Sec.~IV is devoted to explicit development and application of new
classes of parameterized EWs designed to detect genuine tripartite
entanglement in the NIFG.  First, adapting ideas from the work of
G\"uhne \emph{et al.} \cite{Guh1} and V\'ertesi \cite{Vert1}, we
consider $W$ entanglement witnesses based on a periodic spin-chain
model. Second, we introduce a class of $GHZ$ EWs which are
constructed from projection operators corresponding to all the
different types of state vector belonging to a system of three
spin-1/2 fermions. Third, in order to identify the type of genuine
multipartite entanglement that is present, we apply a
stabilizer-operator formalism \cite{Gottes,Toth1,Toth2}. It is
found that $GHZ\setminus W$ genuine tripartite entanglement is not
generated in the ground state of the NIFG as represented by the
three-fermion reduced density matrix, and that genuine $W$-type
tripartite entanglement, although present, does not exist in the
absence of bipartite entangled and/or fully separable states. On
the other hand, some genuine multipartite entangled states that
have a positive partial transpose (PPT) with respect to some
bipartition are found to occur. These conclusions are summarized
in Sec.~VI. Some details relating to the expectation values of the
operators over the most general form of quantum states in the $W$
set as well as the $B$ set are collected in an appendix.

\section{Indistinguishability and the Noninteracting Fermi Gas}

Consider a system consisting of $n$ parties $\{ M_{i}
\}_{i=1}^{n}$. A $k$-partite split is a partition of the system
into $k\leq n$ sets $\{S_{i}\}_{i=1}^{k}$, each of which may be
composed of several original parties.
A given state $\rho\in
\mathcal{B}({\cal{H}}_{d_{1}}\otimes...\otimes{\cal{H}}_{d_{k}})$
associated with some $k$-partite split is called $m$-separable if
a convex decomposition of it can be found such that, in each
pure-state term, at most $m$ parties are mutually entangled, these
not being entangled with any of the other $n-m$ parties.
In particular, a 1-separable ($\equiv$ separable) density matrix
operator $\rho \in {\cal{B}}({\cal{H}})$ (belonging to the Hilbert
space of bounded operators acting on
$\mathcal{H}={\mathcal{H}}_{d_{1}}\otimes...\otimes{\mathcal{H}}_{d_{n}}$)
is fully separable, being expressible as
\begin{equation}
\rho_{\rm{sep}}=\sum_{i} p_{i} | \alpha_{i}^{(1)} \rangle \langle
\alpha_{i}^{(1)} |\otimes | \alpha_{i}^{(2)} \rangle \langle
\alpha_{i}^{(2)} |\otimes...\otimes| \alpha_{i}^{(n)} \rangle
\langle \alpha_{i}^{(n)} |,
\end{equation}
with $p_{i}\geq 0$ and $\sum_{i} p_{i}=1$.
The system is called entangled when the corresponding density
matrix operator is not separable.  According to these definitions,
separable states necessarily form a convex set, since any convex
combination of separable states is again separable---which is not
the case for non-separable states. Beyond bipartite splittings,
many different types of entanglement among the parties are
possible, even for the case of distinguishable particles.

A schematic model involving two electrons located in a double-well
was discussed in Refs.~\onlinecite{Eck1,Ghir1,Am1} to illustrate
the consequences of indistinguishability for entanglement. (See
especially the related treatment of Ref.~\onlinecite{Schl}.) The
qubit is modeled by the spin degree of freedom (with states
$|\uparrow\rangle$ and $|\downarrow\rangle$), and there are two
spatial wave functions labeled $|\phi\rangle$ and $|\chi\rangle$,
initially localized in the left and right potential well,
respectively.  For this bipartite system in a pure state, the
authors considered the density operator $\rho_w=|w\rangle\langle
w| \in {\cal{A}}({{\cal{C}}^{2K} \otimes {\cal{C}}^{2K} })$.
Denoting by $f_{a}$ and $f_{a}^\dag$ the fermionic annihilation
and creation operators for single-particle states $a=1,\ldots\,2K$
constituting an orthonormal basis in ${\cal{C}}^{2K}$, the ket
$|w\rangle$ can be represented as $|w\rangle=\sum_{a,b} w_{ab}
f_a^\dag f_b^\dag|0\rangle$, with the $w_{ab}=-w_{ba}$ defining an
antisymmetric matrix.  For any complex antisymmetric $n\times n$
matrix $(w_{ab})$, there exists a unitary transformation $U$ such
that $w'=U w U^T$ has nonzero entries only in $2\times2$ blocks
along the diagonal \cite{Schl}, i.e.,
\begin{equation}
w'=\mathrm{diag}[ Z_1,...,Z_r,Z_0], \quad Z_{i}=\left(
                                       \begin{array}{cc}
                                         0 & z_i \\
                                         -z_i & 0 \\
                                       \end{array}
                                     \right),
\end{equation}
where $z_i>0$ for $i=1,...,r$ and $Z_0$ is the
$(n-2r)\times(n-2r)$ null matrix. Each $2\times2$ block matrix
$Z_i$ corresponds to an elementary Slater determinant.
The matrix $w'$ enables an expansion of the ket $|w\rangle$ in a
basis of
elementary Slater determinants with a minimum number $r$ of
non-vanishing terms, $r$ being termed the fermionic Slater rank of
$|w\rangle$.
A Slater rank of at least two is required for qualification as an
entangled state.  While this model is illuminating, its extension
to more than two particles becomes very cumbersome, obscuring the
nature of the correlations involved.

An alternative description \cite{Ved1,Oh,Lunk,Vert1}, more fruitful
for our development, places the emphasis on reduced density
matrices of a noninteracting gas of many identical spin-1/2
fermions (NIFG).  By the Pauli exclusion principle, at most two
such particles, with different spin values $s$, can
occupy the same momentum state $p= \hbar k$. The ground state of
the system can be expressed as
\begin{equation}
|\phi_0\rangle = \prod_{s,{\bf p}} b^{\dagger}_s ({\bf p})
|0\rangle , \label{gs}
\end{equation}
where $ [b^{\dagger}_s ({\bf p}), b_{t} ({\bf q})]_{+} =
\delta_{st} \delta ({\bf p}-{\bf q})$ and $| 0 \rangle$ is the
vacuum state.

Although the state of the system is written as product state,
there are specific Pauli-exclusion correlations between the
constituent fermions arising from the commutation rules of the
creation and annihilation operators $b^\dagger$ and $b$.  For the
bipartite and tripartite configurations relevant to our
investigation, these correlations are made explicit by deriving
the two-body and three-body (two-fermion and three-fermion)
reduced density matrices of the pure state (\ref{gs}).
The two-fermion reduced density matrix is given by
\begin{equation}\label{rhoss}
\rho_{ss';tt'} = \langle \phi_0| \psi^{\dagger}_{t'} ({\bf r}')
\psi^{\dagger}_{t} ({\bf r}) \psi_{s'} ({\bf r}') \psi_{s} ({\bf
r}) |\phi_0\rangle ,
\end{equation}
where $\psi^{\dagger}_{s} (r)$ [$\psi_{s} (r)$] creates [destroys]
a particle with spin $s$ at the location ${\bf r}$.  With the
transformation
\begin{equation}\label{psis}
\psi_s ({\bf r}) = \int \frac{d^3 k}{(2\pi)^3} e^{i{\bf
k}\cdot{\bf r}} b_s ({\bf k}) ,
\end{equation}
it is straightforward to obtain the two-body reduced density
matrix in the form
\begin{equation}
\rho_{ss';tt'} = n^2 [\delta_{st}\delta_{s't'} -
\delta_{s't}\delta_{st'} f^2(|\textbf{r}-\textbf{r}'|)] ,
\label{twoelect}
\end{equation}
%
where $n =  k_F^3/6\pi^2$ is the density of particle of a given
spin, and
\begin{equation}
f(|\textbf{r}-\textbf{r}'|) = \int_{0}^{k_F} \frac{d^3
k}{(2\pi)^3} e^{-i{\bf k} \cdot (\bf{r}-\bf{r}'|)}
\end{equation}
is known as the Slater factor.
(We focus here on the NIFG in three dimensions.)
%

To evaluate the bipartite entanglement corresponding to the
two-fermion reduced density matrix, one performs a partial
transposition and determines the eigenvalues of the density matrix
and its partial transpose \cite{Ved1}.  Entanglement exists for
two-fermion configurations such that $f^2>1/2$, i.e., for $0\le
|\textbf{r}-\textbf{r}'| < r_e$, where $r_e $ is cut-off radius
for entanglement (rather than the classical correlation).  For the
3D NIFG, $r_e$ is determined by $j^2_1 (k_F r_e) =1/2$, where
$j_1$ is the first-order spherical Bessel function.

The same steps may be used to derive the reduced three-fermion
density matrix of the NIFG as a function of particle locations
$\textbf{r},\textbf{r}',\textbf{r}''$ and spins $s, s', s''$. Six
possible arrangements give rise to six terms:

\begin{widetext}
\begin{align}\label{fff}
\rho (s,s',s'';t,t',t'')  =& \langle \phi_0|\psi^{\dagger}_{t''}
(r'') \psi^{\dagger}_{t'} (r') \psi^{\dagger}_{t} (r) \psi_{s}
(r') \psi_{s'} (r) \psi_{s''} (r'')
|\phi_0\rangle \nonumber \\
=&   n^3 (\delta_{st}\delta_{s't'}\delta_{s''t''} -  f_{12} f_{13}
\delta_{st}\delta_{s't''}\delta_{s't''} - f_{13} f_{23}
\delta_{st'}\delta_{s't}\delta_{s''t''} \nonumber \\
&- f_{12} f_{23} \delta_{st''}\delta_{s't'}\delta_{s''t}  + f_{12}
f_{13} f_{23} \delta_{st'}\delta_{s't''}\delta_{s''t} +  f_{12}
f_{13} f_{23} \delta_{st''}\delta_{s't}\delta_{s''t'}) .
\end{align}
\end{widetext}
The three functions $f_{12}$, $f_{13}$, and $f_{23}$ carry the
respective arguments $|\textbf{r}-\textbf{r}'|$,
$|\textbf{r}-\textbf{r}''|$, and $|\textbf{r}'-\textbf{r}''|$.

The $9 \times 9$ matrix defined by Eq.~(\ref{fff}) takes the form
\begin{widetext}
\begin{eqnarray}
\label{rho3matrix}
\rho_{3}=\left(\begin{array}{cccccccc}
\eta&0&0&0&0&0&0&0\\
0&\eta+\frac{p_{13}+p_{23}}{4}&\frac{-p_{23}}{4}&0&\frac{-p_{13}}{4}&0&0&0\\
0&\frac{-p_{23}}{4}&\eta+\frac{p_{12}+p_{23}}{4}&0&\frac{-p_{12}}{4}&0&0&0\\
0&0&0&\eta+\frac{p_{12}+p_{13}}{4}&0&\frac{-p_{12}}{4}&\frac{-p_{13}}{4}&0\\
0&\frac{-p_{13}}{4}&\frac{-p_{12}}{4}&0&\eta+\frac{p_{13}+p_{12}}{4}&0&0&0\\
0&0&0&\frac{-p_{12}}{4}&0&\eta+\frac{p_{23}+p_{12}}{4}&\frac{-p_{23}}{4}&0\\
0&0&0&\frac{-p_{13}}{4}&0&\frac{-p_{23}}{4}&\eta+\frac{p_{23}+p_{13}}{4}&0\\
0&0&0&0&0&0&0&\eta\end{array} \right) ,
\end{eqnarray}
\end{widetext}
where $\eta=(1-p_{12}-p_{13}-p_{23})/8$ and
\begin{equation}\label{pijs}
p_{ij}=\frac{-f^2_{ij}+f_{ij}f_{ik}f_{jk}}{-2+f^2_{ij}+f^2_{ik}+f^2_{jk}-f_{ij}f_{ik}f_{jk}}.
\end{equation}

Lunkes \emph{et al.}~\cite{Lunk} have provided the following
general expression for the $n$-body reduced density matrix of the
NIFG,
\begin{equation}\label{eq:dm}
\rho_{n}=(1-\sum_{ij}p_{ij})\frac{\textbf{I}}{2^n}
+\frac{1}{2}\sum_{ij, i \neq j}p_{ij}|\Psi_{ij}^{-}
\rangle\langle\Psi_{ij}^{-}|\otimes \frac{\textbf{I}}{2^{n-2}},
\end{equation}
which is constructed from biseparable entangled density operators.
Here, $|\Psi^{-}_{ij}\rangle =2^{-1/2}(|01\rangle-|10\rangle)$ is
the maximally entangled singlet state of the pair $ij$, and
$|f_{ij}|\leq 1$ for all $ij$.  Consequently $|p_{ij}|\leq 1$ and
$0\leq\eta\leq\frac{1}{4}$, so we have
$|p_{12}+p_{13}+p_{23}|\leq1$ \cite{Vert1}. This indicates that
the density matrix of $n$ noninteracting fermions can be written
in terms of antisymmetric density matrices of fermionic pairs,
albeit not in a convex combination.  Considering the
aforementioned entanglement condition for two-fermion
configurations, entanglement would be present when all $f^2$
factors are greater than $1/2$.  Despite the explicit form
(\ref{eq:dm}) of the $n$-body reduced density matrix of the NIFG
as a combination of biseparable states, the existence of genuine
tripartite entanglement in this system was established in
Ref.~\onlinecite{Vert1}.

The following sections will build upon this
important result.  We shall formulate new classes of three-qubit
entanglement witnesses (EWs) and demonstrate that they can detect
the corresponding fermionic tripartite density matrix
\begin{align}\label{NIFG}
\rho_3=&\eta\textbf{I}
+a|\Psi_{12}^{-}\rangle\langle\Psi_{12}^{-}|\otimes
\frac{\textbf{I}}{2}+b|\Psi_{13}^{-}\rangle\langle\Psi_{13}^{-}|\otimes
\frac{\textbf{I}}{2}\nn\\&+c|\Psi_{23}^{-}\rangle\langle\Psi_{23}^{-}|\otimes
\frac{\textbf{I}}{2}\nn\\
=&\frac{1}{8}\textbf{I}-\frac{a}{8}(\sigma_x \sigma_x
\textbf{I}+\sigma_y \sigma_y \textbf{I}+\sigma_z \sigma_z
\textbf{I})\nn\\&-\frac{b}{8}(\sigma_x \textbf{I}\sigma_x
+\sigma_y \textbf{I}\sigma_y +\sigma_z \textbf{I}\sigma_z
)\nn\\&-\frac{c}{8}(\textbf{I}\sigma_x \sigma_x
+\textbf{I}\sigma_y \sigma_y + \textbf{I}\sigma_z \sigma_z)
\end{align}
of the NIFG. (It is convenient to relabel the quantities $p_{ij}$
of Eq.~(\ref{pijs}) through $a=p_{12}$, $b=p_{13}$, and
$c=p_{23}$.) Obviously, $\rho_3$ cannot possess genuine tripartite
entanglement for the case of simultaneous positive values of $a$,
$b$, and $c$, due to the definition of biseparable states.

\begin{figure}[]
\begin{center}
\subfigure[]{
\includegraphics[width=6cm]{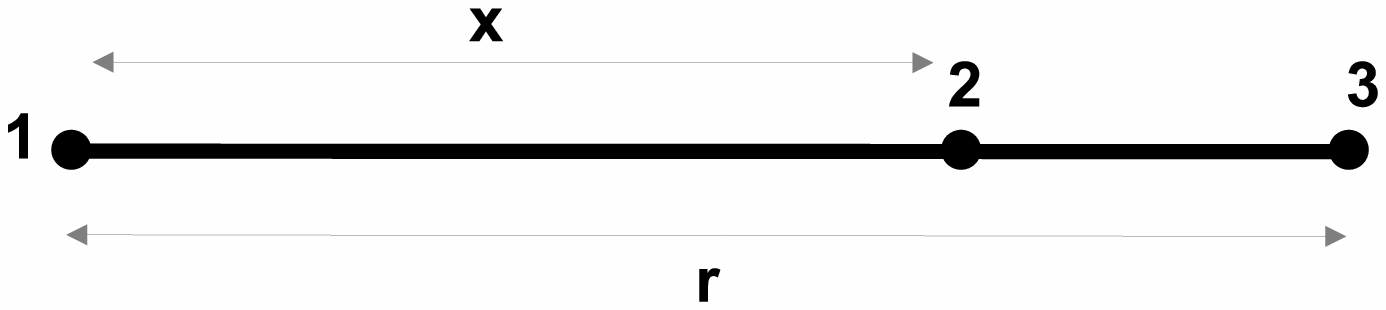}}
\subfigure[]{
\includegraphics[width=8.5cm]{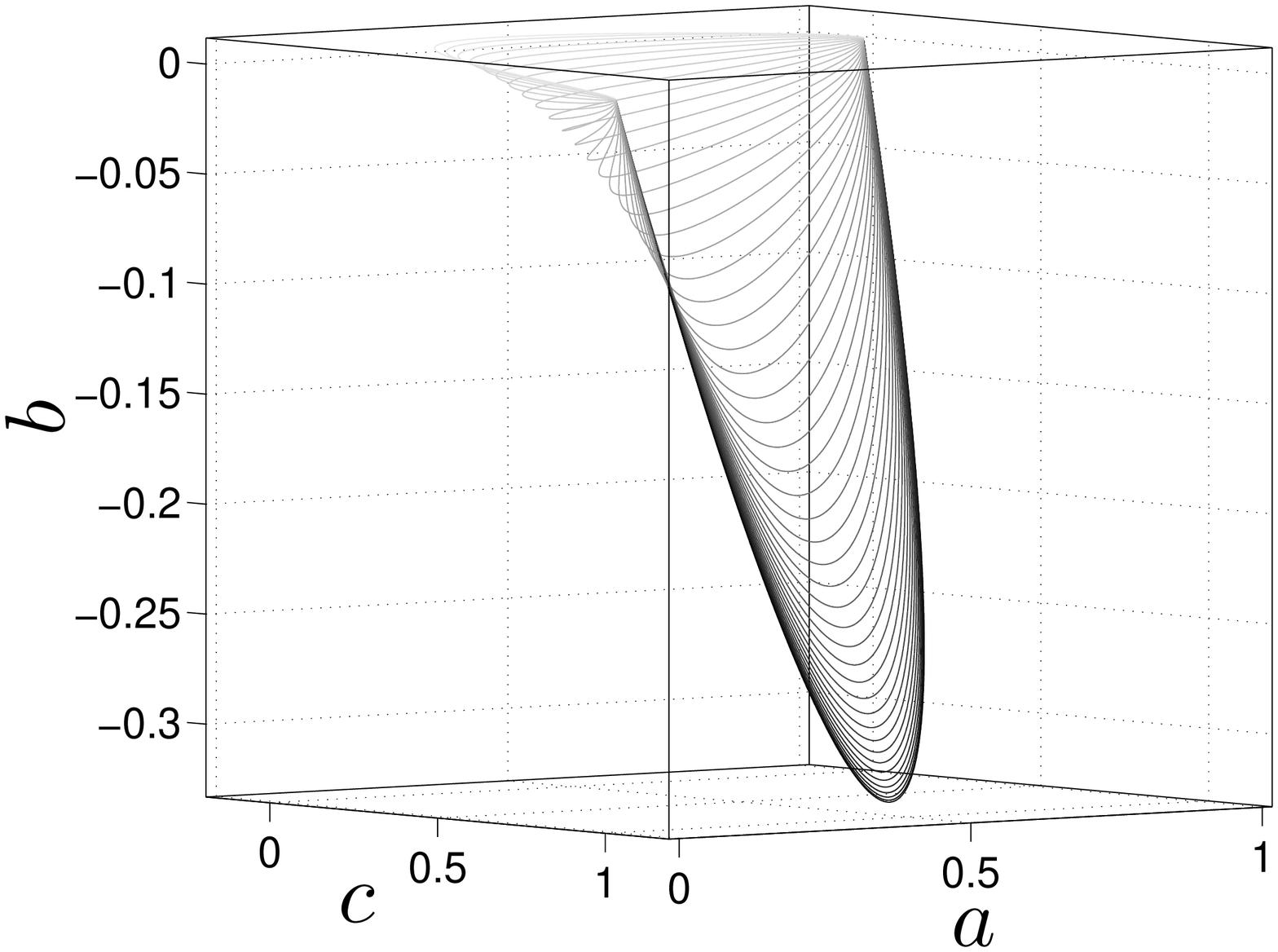}}
\end{center}
\caption[]{For 1-d configurations as specified in (a), the
coefficients $a$ and $b$ are plotted versus $c$ in (b) at
different values of the quantity $k_Fx$. Plot traces change from
black to light gray as $k_Fx$ increases from 0.1 to 5.0 in steps
of 0.1. For $k_Fr\geq4.5$, three $p$ functions attain positive
values, placing the configuration outside the regime of genuine
tripartite entanglement.}\label{p_1d}
\end{figure}

\begin{figure}[]
\begin{center}
\subfigure[]{
\includegraphics[width=7cm]{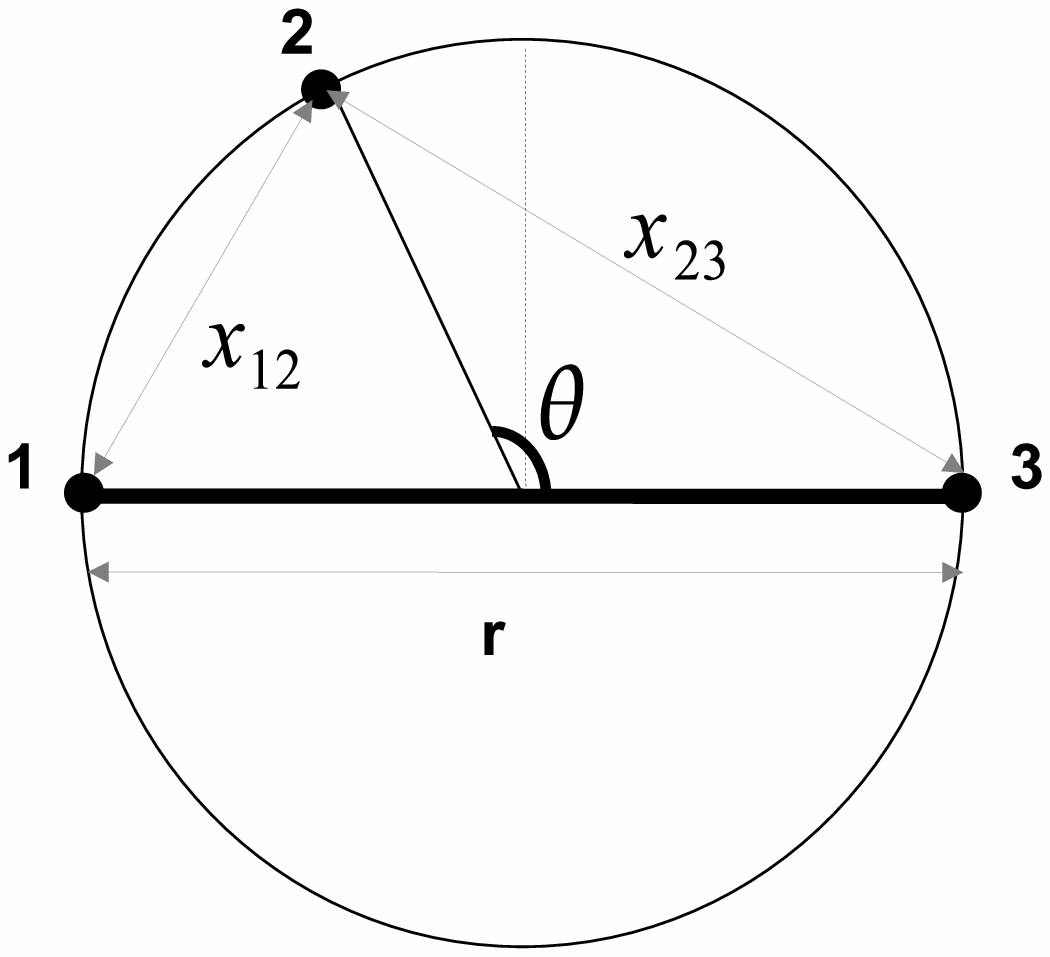}}
\subfigure[]{
\includegraphics[width=8cm]{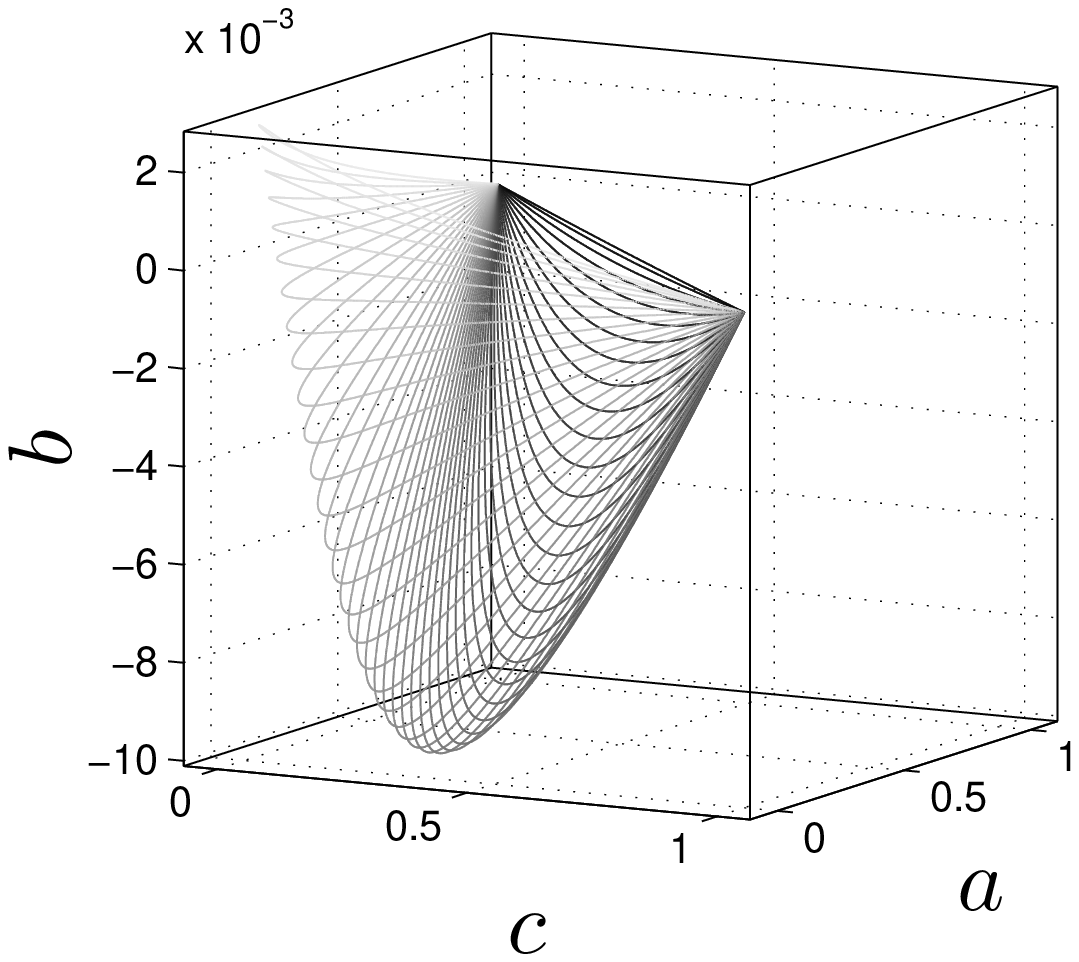}}
\end{center}
\caption[]{
For 2-d configurations as specified in (a), the coefficients $a$
and $b$ are plotted versus $c$ in (b) at different values of $k_Fx$.
Particle 2 is considered to move on a circle of fixed radius $r$,
with $x_{12}=r\cos{(\theta/2)}$ and $x_{23}=r\sin{(\theta/2)}$.
The behavior of the coefficients is shown with traces changing
from black to light gray as $k_F r$ increases from 0.1 to 5.0 in
steps of 0.1.} \label{p_2d}
\end{figure}

It is instructive to sample the behavior of the coefficients $a$,
$b$, and $c$ of Eq.~(\ref{NIFG}) for the NIFG.  For
one-dimensional (1-d) configurations specified by the distance $x$
between fermions 1 and 2 and the distance $r$ between 1 and 3,
Fig.~\ref{p_1d} shows a plot of $b$ with respect to $a$ and $c$
under varying $k_Fr$.  Fig.~\ref{p_2d} provides a similar plot for
two-dimensional configurations in which fermions 1 and 3 are
separated by $r$ and fermion 2 is constrained to move on a circle
of radius $r$ centered midway between 1 and 3.  It is found that
genuine tripartite entanglement cannot be generated in the region
$4.5\leq k_Fr\leq5$.

Quite apart from consideration of the
entanglement properties specific to the NIFG,
the coefficients $a$, $b$, and $c$ in Eq.~(\ref{NIFG}) must
satisfy the inequalities
\begin{align}\label{constdens}
  &\frac{1}{8}+\frac{1}{8}(a+b+c)
  \pm \frac{1}{4}  \sqrt{a^2+b^2+c^2-ab-bc-ac}\geq 0 , \nn\\
  & a^2+b^2+c^2\geq ab+bc+ac , \nn\\
  & \eta\geq 0 ,
\end{align}
imposed by the restriction of the eigenvalues of any density
operator to positive-definite values.  Referring to the
definition of biseparable states, we note that by the definition
of biseparable states, only negative values of the coefficients
$a$, $b$, and $c$ can give rise to genuine tripartite entanglement
in $\rho_3$.

\section{Entanglement Witnesses}

The existence of an entanglement witness (EW) for any type of
entangled state follows from the Hahn-Banach theorem
\cite{rudin1}. In essence, this theorem establishes that if $C_1$
and $C_2$ are convex closed sets in a real Banach space, one of
them being compact, there exists a bounded functional (identified
here as the witness operator) that serves to separate the two
sets.  For example, an EW can be defined for entanglement class
$e$ as an Hermitian operator $\mathcal{W}$ such that (a)
$\mathrm{Tr}(\mathcal{W}\rho_{s})\geq 0 $ for all fully separable
states $\rho_{s}$ and (b) there exists at least one entangled
state $\rho_{e}$ which can be detected by the condition
$\mathrm{Tr}(\mathcal{W}\rho_{e})<0 $. It will be relevant to
later developments that this definition in itself cannot
distinguish between different kinds of entanglement for more than
bipartite systems.

The objective of the present work is the elucidation of the
tripartite entanglement properties of the NIFG.  This objective is
pursued within the framework of the comprehensive analysis and
classification of mixed three-qubit states provided by Ac\'in
\emph{et al.}\ \cite{Acin}.  This classification is a
generalization of that for pure three-qubit states.  To introduce
the necessary definitions and results, we first recall that the
most general pure three-qubit state takes the form
\begin{align}\label{ghz}
|\psi_{GHZ}\rangle=&\lambda_0|000\rangle+\lambda_1
e^{i\theta}|100\rangle+\lambda_2|101\rangle\nn\\&
+\lambda_3|110\rangle+\lambda_4|111\rangle,
\end{align}
with $\lambda_i \geq 0$, $\sum_i \lambda_i^2 = 1$, and $\theta\in
[0,\pi]$.  For the three-qubit system, there are two types of
locally inequivalent entangled pure states.  These are the $GHZ$
type, defined by the form (\ref{ghz}) with nontrivial parameters
(especially, $\lambda_4 > 0$), and the $W$ type, represented
generically by
\begin{equation}\label{w}
|\psi_W\rangle=\lambda_0|000\rangle+\lambda_1|100\rangle+\lambda_2|101\rangle+\lambda_3|110\rangle.
\end{equation}
The analysis of Ref.~\onlinecite{Acin} identifies four distinct classes
of mixed three-qubit states.  The corresponding sets, denoted by
$S$, $B$, $W$, and $GHZ$, are all convex and compact, each being formed
as a convex sum of appropriate projectors.  Specifically,
$S$ states are mixtures of product vectors; $B$ states are mixtures
of product and biseparable vectors; $W$ states in turn are mixtures
of product vectors, biseparable states, and $W$ vectors (\ref{w});
and $GHZ$ states are mixtures of all of the previous vector types as well
as $GHZ$ vectors (\ref{w}).  These four sets are invariant under
local unitary and invertible non-unitary transformations. Evidently,
they are nested according to $S \subset B \subset W \subset GHZ$.

Our development will also involve non-convex subsets such as
$B\setminus S$, $W\setminus S$, $W\setminus B$, $GHZ\setminus S$,
$GHZ\setminus B$, and $GHZ\setminus W$, which exclude fully
separable states and hence contain only entangled states.   A
three-qubit state is said to possess genuine tripartite
entanglement if it does not belong to the class of biseparable
states $B$; such a state resides either in $GHZ\setminus W$ or
$W\setminus B$.

The classification of mixed 3-qubit states introduced by Ac\'in et
al.\ \cite{Acin} allows for the construction of entanglement
witnesses---namely $GHZ$ EWs and $W$ EWs---that are capable of
detecting states with genuine tripartite entanglement.  Thus, $W$
EW will denote an operator $\mathcal{W}_W$ such that
$\mathrm{Tr}(\mathcal{W}_W \rho_B)\geq 0$ holds $\forall \rho_B\in
B$, but for which there exists a $\rho_{GHZ\setminus B} \in
GHZ\setminus B$ such that $\mathrm{Tr}(\mathcal{W}_W
\rho_{GHZ\setminus B})<0$, thereby discriminating between the sets
$B$ and $GHZ\setminus B$.  Similarly, a $GHZ$ EW is provided by an
operator $\mathcal{W}_{GHZ}$ such that
$\mathrm{Tr}(\mathcal{W}_{GHZ} \rho_W)\geq 0$ is satisfied for any
$W$ state, but produces a negative expectation value for some
$GHZ\setminus W$ state.  Additionally, genuine tripartite
entangled states, belonging to $GHZ\setminus W$ and $W\setminus
B$, can be identified by means of an EW operator designed to
detect $W\setminus B$ entanglement---as will be demonstrated in
the next section.

A general scheme utilizing linear programming (LP) to arrive at EW
operators for the detection of $GHZ\setminus S$ entangled states
was introduced in Refs.~\onlinecite{hes1,hes2}. To confirm the
presence of genuine tripartite entanglement in the NIFG one needs
$W$ or $GHZ$ EWs.  Construction of parameterized $GHZ$ EWs via the
linear programming algorithm proceeds as follows. First, consider
an Hermitian operator $\mathcal{W}$ in the form
\begin{equation}\label{witdef}
 \mathcal{W}=\sum_{i}a_{i} \widehat{P}_{i},
\end{equation}
possessing at least one negative eigenvalue, where the
$\widehat{P}_{i}$'s are Hermitian linear operators such that $x
\leq \mathrm{Tr}(\widehat{P}_{i} \rho_{W})\leq y$ for all $x,y\in
\mathbb{R}$ and for any density operator $ \rho_{W}\in W$. The
parameters $a_{i}\in \mathbb{R}$ are to be determined such that
$\mathcal{W}$ qualifies as a $GHZ$ EW.  As $\rho_W$ varies over
$W$ states, $P^W_i=\mathrm{Tr}(\widehat{P}_i\rho_W)$ maps $W$
states into a convex region, since a linear functional maps a
convex domain (here, the $W$ class) to a convex region.  Our
principal task is to choose proper operators $\widehat{P}_i$ so as
to obtain an approximating convex polyhedron surrounding the
convex region spanned by the $P^W_i$ (the so-called feasible
region for the LP optimization). The term ``approximating'' refers
to the fact that in general, not all points in the convex
polyhedron are produced by the expectation values of the
$\widehat{P}_{i}$'s over $W$ states. By using this approximating
convex polyhedron, the expectation value of the operator
$\mathcal{W}$ over all $W$ states is non-negative and, with
respect to other states, admits at least one negative value.
Accordingly, in seeking to determine a $GHZ$ EWs of type
(\ref{witdef}), one needs to find the minimum expectation value of
$\sum_{i}a_iP^W_i$ over the feasible region. In this way, the
problem is reduced to optimization of the linear function
$\sum_{i}a_iP^W_i$ over the convex set provided by the
approximating convex polyhedron.

In characterizing the properties of the density operator for the
NIFG, it is of interest to investigate the possibility of
positive-partial transpose (PPT) entanglement, specifically the
presence of a PPT entangled state with respect to some bipartition
of a three-particle subsystem. The decomposability or
non-decomposability of an EW may depend on which particles are
involved. By definition, an EW $\mathcal{W}$ is partially
decomposable with respect to the $i$-th party iff there exist
positive operators $\widehat{P}, \widehat{Q}$ such that
$\mathcal{W}=\widehat{P}+\widehat{Q}^{T_{i}}$, where
$i\in\{1,2,3\}$ and $T_i$ stands for partial transposition with
respect to $i$-th party. Accordingly, an EW is called partially
non-decomposable EW with respect to a given party iff there exists
at least one PPT entangled state associated with that party, i.e.,
$\rho^{T_{i}}\geq0$ when $ \mathrm{Tr}( \mathcal{W} \rho )<0$. The
relevance of these definitions will become clear when we encounter
EWs that must explicitly bear the label of the party (or particle)
with respect to which it is decomposable.  For the purpose of
identifying a PPT state with respect to a given party, one may
derive  the following conditions for the parameters in $\rho_3$: A
positive value for $\rho^{T_1}$ requires
\begin{align}\label{cond1}
&a^2+b^2+c^2+ac+bc-ab\geq0, \nn\\
&2(a^2+b^2+c^2+ac+bc-ab)^{1/2}-1\leq -a-b+c \leq 1,
\end{align}
the condition $\rho^{T_2}\geq0$ is met only if
\begin{align}
&a^2+b^2+c^2-ac+bc+ab\geq0,\nn\\
&2(a^2+b^2+c^2-ac+bc+ab)^{1/2}-1\leq -a+b-c \leq 1, \label{cond2}
\end{align}
and $\rho^{T_3}\geq0$ implies
\begin{align}
&a^2+b^2+c^2+ac-bc+ab\geq0,\nn\\
&2(a^2+b^2+c^2+ac-bc+ab)^{1/2}-1\leq a-b-c \leq 1. \label{cond3}
\end{align}
The presence of a PPT entangled density matrix with respect to a
given party is signaled by satisfaction the corresponding
necessary condition among (\ref{cond1}-\ref{cond3}) together with
its detection by the corresponding EW operator. Evidently, only a
partially non-decomposable EW related to the selected party can
detect a PPT entangled state with respect to that party.

\section{EW operators for the NIFG}

In accordance with the strategy proposed in
Ref.~\onlinecite{hes2}, we now apply the LP method to determine
proper EW operators for identification of chosen entanglement
classes in the density matrix of the NIFG.  We first consider EW
construction based on an extended spin-chain model, and then build
a new class of $GHZ$ EWs from an appropriate set of density
operators.  Finally, we utilize stabilizer operators to develop
EWs that discriminate between different classes of genuine
tripartite entanglement.

\subsection{Spin-chain model}\label{subsecA}
G\"{u}hne \emph{et al.} \cite{Guh1} were the first to introduce an
EW operator ($W$ EW) to detect genuine tripartite entanglement
based on a macroscopic spin-chain model.  The explicit form of
this EW is
\begin{align}\label{guhne}
\mathcal{W}^{({\rm gen})}=(1+\sqrt{5})\textbf{I}+
\widehat{P}_{12}+ \widehat{P}_{23},
\end{align}
where $\widehat{P}_{ij}=\sigma^{(i)} \cdot\sigma^{(j)}$, or more
rigorously
\begin{align}\label{Pij}
\widehat{P}_{12}=&\sigma^{(1)} \cdot\sigma^{(2)}
\nn\\=&\sigma_x^{(1)}\sigma_x^{(2)}
\textbf{I}^{(3)}+\sigma_y^{(1)}\sigma_y^{(2)} \textbf{I}^{(3)}
+\sigma_z^{(1)} \sigma_z^{(2)}\textbf{I}^{(3)}\nn,\\
\widehat{P}_{23}=&\sigma^{(2)}
\cdot\sigma^{(3)}\nn\\=&\textbf{I}^{(1)}\sigma_x^{(2)}
\sigma_x^{(3)} +\textbf{I}^{(1)}\sigma_y^{(2)} \sigma_y^{(3)}
+\textbf{I}^{(1)}\sigma_z^{(2)} \sigma_z^{(3)}.
\end{align}
The superscripts identify the parties involved and
$\sigma=(\sigma_x,\sigma_y,\sigma_z)$. As required, the operator
$\mathcal{W}^{({\rm gen})}$ has positive expectation values with
respect to all states belonging to $B$ and has a negative
expectation value with respect to some genuine tripartite
entangled state in $GHZ\setminus B$.

Implementing the EW of G\"uhne \emph{et al.}\ \cite{Guh1},
V\'{e}rtesi \cite{Vert1} has shown that in a particular
three-fermion configuration, there exists genuine tripartite
entanglement in the NIFG.  Our goal is to identify the {\it type}
of genuine tripartite entanglement present in the NIFG.
We develop EW operators suitable for this system and seek
boundary conditions for separated fermions exploiting the witness
operators.

Starting from spin-chain model used in Ref.~\onlinecite{Guh1}
to construct the EW of Eq.~(\ref{guhne}), we consider the
operator
\begin{equation}\label{guhne1}
\mathcal{W}_{1}=c_0\;\textbf{I}+ \widehat{P}_{12}+
\widehat{P}_{23}
\end{equation}
and ask whether the constant $c_0$ can be chosen so that
$\mathcal{W}_{1}$ becomes a $GHZ$ EW.  To answer this question, we
should find the minimum expectation value of $\widehat{P}_{12}+
\widehat{P}_{23}$ over all states in $W$ set. The domain of $W$
states is spanned by the explicit general form (\ref{w}) for
$|\psi_W\rangle$, subject to any local $SU(2)$ transformation.
Upon expanding such the general $W$ vector as $|\Psi_W\rangle=
\sum_{i,j,k=0,1}A_{ijk}|ijk\rangle$, one can evaluate the
coefficient $A_{ijk}$ and thus the expectation values $P_{ij}^W =
\langle \Psi_W| P_{ij} |\Psi_W\rangle$.  (See the appendix for
details.) It is then easily seen that the witness
$\mathcal{W}_{1}$ can only detect $W\setminus B$ states, since the
eigenvector associated with the minimum eigenvalue $-4$ of
$\widehat{P}_{12}+ \widehat{P}_{23}$ belongs to $W$ domain, so
$\mathcal{W}_1$ can never be a $GHZ$ EW.

Evaluating the expectation value of (\ref{guhne}) with respect to
$\rho_3$, we obtain
\begin{equation}
\mathrm{Tr}(\mathcal{W}^{({\rm gen})}\rho_3)=1+\sqrt{5}-3(a+c),
\end{equation}
which cannot be negative and maintain the condition $a+b+c<1$
specific to the NIFG.

\begin{figure}[]
\begin{center}
\includegraphics[width=7.5cm]{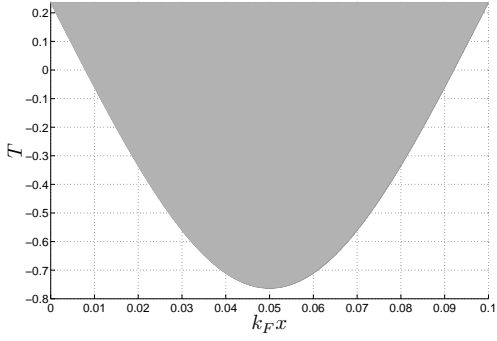}
\end{center}
\caption[]{Plot of the expectation value
$T=\mathrm{Tr}(\mathcal{W}^{({\rm gen})}\rho_3')$ versus $k_Fx$ at
$k_Fr=0.1$, in the 1-d coordinate scheme of Fig.~\ref{p_1d}. Here,
$\mathcal{W}^{({\rm gen})}$ is the entanglement witness
(\ref{guhne}) and $\rho_3'$ is the transform (\ref{rho3prime}) of
the three-fermion reduced density matrix $\rho_3$ under an
arbitrary local unitary operator.  Physically, the distance of
particle 2 from particle 1 is varied while the distance between
particles 1 and 3 is kept fixed.  The trace $T$ lies inside the
gray region when $k_Fx$ varies between 0 and $k_F r$ and the
parameters $\alpha$ and $\beta$ range over all allowed values.
}\label{P12P23}
\end{figure}

To investigate the existence of genuine tripartite entanglement,
we adopt the more general form
\begin{equation}\label{rho3prime}
\rho_3'=U\rho_3 U^{\dag}
\end{equation}
for the density operator, where
\begin{equation}\label{Udef}
U=\left(
\begin{array}{cc}
    \beta^* & \alpha \\
    -\alpha^* & \beta \\
\end{array}
\right)^{\otimes 3}
\end{equation}
performs an arbitrary local unitary transformation of the density
operator $\rho_3$ of the NIFG.  (The same transformation is
applied for all three fermions, with $|\alpha|^2+|\beta|^2=1$.) We
then find
\begin{align}\label{}
\mathrm{Tr}(\mathcal{W}^{({\rm gen})}\rho_3')
=1+\sqrt{5}+(2-2b-7a-7c)|\beta^2\alpha^4|\nn\\
-9(a+c)|\alpha^2\beta^4|- (a+c)|\alpha|^6-3(a+c)|\beta|^6.
\end{align}
For one-dimensional configurations specified as in
Fig.~\ref{p_1d}, Fig.~\ref{P12P23} shows the value of
$\mathrm{Tr}(\mathcal{W}^{({\rm gen})}\rho_3')$ as a function of
$k_Fx$ for $k_Fr=0.1$.  It is seen that $W\setminus B$
entanglement is present over the range $0.01\leq k_Fx\leq0.09$.

Acknowledging the indistinguishability of the fermion constituents
of the NIFG, we turn to a more general EW construction based on a
parametrized operator that superposes all three of the spin
products $\widehat{P}_{ij}$:
\begin{equation}\label{Wspin1}
\mathcal{W}^{(\mathrm{sp})}=a_0\; \textbf{I}+ a_{12}
\widehat{P}_{12}+ a_{13} \widehat{P}_{13}+ a_{23}
\widehat{P}_{23}.
\end{equation}
For equal values of the real parameters $a_{12}=a_{13}=a_{23}$,
$\mathcal{W}^{(\mathrm{sp})}$, this ansatz reduces to
$\mathcal{W}^{(\mathrm{sp})}=a_0
\mathbf{I}_8+a_{12}(\widehat{P}_{12}+\widehat{P}_{13}
+\widehat{P}_{23})$ which resembles a spin-chain model with
periodic boundary condition.  We now consider the possibility that
$\mathcal{W}^{(\mathrm{sp})}$ can provide a $W$ EW or even a $GHZ$
EW by examining the constraints on the parameters $a_0$, $a_{12}$,
$a_{13}$, and $a_{23}$.

In establishing $\mathcal{W}^{(\mathrm{sp})}$ as a $W$ EW
detecting genuine tripartite entanglement, the essential task is
to find a convex polyhedron spanned by the $P^B_{ij}$, i.e., the
expectation values of the $\widehat{P}_{ij}$ with respect to the
$B$ class of states (as considered in the appendix). As a first
conjecture delimiting the eigenvalues of the $\widehat{P}_{ij}$,
we propose the set of inequalities
\begin{align}\label{poly1}
-3\leq P^B_{12},P^B_{13},P^B_{23}\leq1\nn,\\
-P^B_{12}-P^B_{13}+P^B_{23}\leq3\nn,\\
-P^B_{12}+P^B_{13}-P^B_{23}\leq3\nn,\\
P^B_{12}-P^B_{13}-P^B_{23}\leq3.
\end{align}
The polyhedron so described does not encompass the region spanned
by the $P^B_{ij}$. However, by parallel shifts of the boundaries
in relations (\ref{poly1}), one can find a proper approximating
convex polyhedron for reduction of the problem of finding a $W$ EW
to one of linear programming.  This step is outlined in the
appendix.   The resulting optimization problem reads:
\begin{align}\label{lp3}
&\mathrm{Minimize} \quad\;\; a_0+a_{12} P^B_{12}+a_{13} P^B_{13}
+a_{23} P^B_{23},\nonumber\\
&\mathrm{subject\ to} \quad\;\;\left\{\begin{array}{c}
  -3\leq P^B_{12},P^B_{13},P^B_{23}\leq1,\\
  -P^B_{12}-P^B_{13}+P^B_{23}\leq 1+\sqrt{8},\\
  -P^B_{12}+P^B_{13}-P^B_{23}\leq 1+\sqrt{8},\\
  P^B_{12}-P^B_{13}-P^B_{23}\leq 1+\sqrt{8}.
\end{array}\right.
\end{align}
With regard to maximum eigenvalues, the expectation values of
\begin{align}\label{operators}
\widehat{P}_{23}-\widehat{P}_{12}-\widehat{P}_{13},\nn\\
\widehat{P}_{13}-\widehat{P}_{12}- \widehat{P}_{23},\nn\\
\widehat{P}_{12}-\widehat{P}_{13}- \widehat{P}_{23}
\end{align}
reach 5.  Therefore, solution of the LP problem (\ref{lp3}) determines a
region in which to form $\mathcal{W}^{(\mathrm{sp})}$ as a witness
operator.  Solution proceeds by considering the intersection of the
constraints expressed in (\ref{lp3}) and finding the vertices of the convex
polyhedron.  Upon solving the LP problem, one can obtain the values
of the parameters $a_{0}$, $a_{12}$, $a_{13}$ and $a_{23}$ satisfying
\begin{align}\label{aij}
&a_0+(3-\sqrt{8})a_{12}-3a_{13}+a_{23}\geq0\nn,\\
&a_0-3a_{12}+(3-\sqrt{8})a_{13}+a_{23}\geq0\nn,\\
&a_0-3a_{12}-3a_{13}+(-5+\sqrt{8})a_{23}\geq0\nn,\\
&a_0+(3-\sqrt{8})a_{12}+a_{13}-3a_{23}\geq0\nn,\\
&a_0-3a_{12}+a_{13}+(3-\sqrt{8})a_{23}\geq0\nn,\\
&a_0-3a_{12}+(-5+\sqrt{8})a_{13}-3a_{23}\geq0\nn,\\
&a_0+a_{12}-3a_{13}+(3-\sqrt{8})a_{23}\geq0\nn,\\
&a_0+a_{12}+(3-\sqrt{8})a_{13}-3a_{23}\geq0\nn,\\
&a_0+(-5+\sqrt{8})a_{12}-3a_{13}-3a_{23}\geq0\nn,\\
&a_0+a_{12}+a_{13}+a_{23}\geq0\nn,\\
&a_0+a_{12}+a_{13}-3a_{23}\geq0\nn,\\
&a_0-3a_{12}+a_{13}+a_{23}\geq0\nn,\\
&a_0+a_{12}-3a_{13}+a_{23}\geq0\nn,\\
&a_0-3a_{12}-3a_{13}-3a_{23}\geq0,
\end{align}
provided that the expectation value of
$\mathcal{W}^{(\mathrm{sp})}$ is positive for all quantum states
in the $W$ set. Additionally, for $\mathcal{W}^{(\mathrm{sp})}$ to
qualify as a $W$ EW, at least one of the eigenvalues $E_i$ among
\begin{align}
E_1&=a_0+a_{12}+a_{13}+a_{23},\nonumber\\
E_{2,3}&=a_0-a_{12}-a_{13}-a_{23}\nn\\
&\pm(a_{12}^2+a_{13}^2+a_{23}^2-a_{12}a_{13}-a_{12}a_{23}-a_{13}a_{23})^{1/2},
\end{align}
must be negative, i.e., $\mathrm{min}(E_i)<0$ for $i=1,2,3$, thus
imposing on the parameters $a_i$ a condition additional to those
of Eq.~(\ref{aij}).

In order to distinguish between different classes of genuine
tripartite entanglement, one would like to find a set of
parameters such that $\mathcal{W}^{(\mathrm{sp})}$ becomes a $GHZ$
EW. Following the same pattern as above, we seek a proper
polyhedron spanned by the $P^W_{ij}$ (see appendix, Eq.~(A-5)). It
is found that the maximum value of $-P^W_{12}-P^W_{13}+P^W_{23}$,
$-P^W_{12}+P^W_{13}-P^W_{23}$, and $P^W_{12}-P^W_{13}-P^W_{23}$
reaches $5$, which is the maximum eigenvalue of the operators in
Eqs.~(\ref{operators}). Hence $\mathcal{W}^{(\mathrm{sp})}$ can
never be a $GHZ$ EW for any choice of the parameters $a_i$. It may
be noted that use of the form $\mathcal{W}^{(\mathrm{sp})}$ in
construction of an EW operator is compatible with the idea of
developing it in terms of the projection operators
$|\Psi_{ij}^{-}\rangle\langle\Psi_{ij}^{-}|\otimes
\textbf{I}^{(k)}$ 
present in the expansion (\ref{NIFG}) of $\rho_3$,
since one can write
\begin{equation}\label{spinproj}
|\Psi_{ij}^{-}\rangle\langle\Psi_{ij}^{-}|\otimes
\textbf{I}^{(k)}=\frac{1}{4}(\textbf{I}-\sigma^{(i)}
\cdot\sigma^{(j)}).
\end{equation}
However, the more general (and more successful) implementation of
this idea pursued in Sec.~\ref{subsecB} takes into account
projectors referring to the set $GHZ\setminus B$.

Returning to the task of detecting genuine tripartite entanglement
in the three-particle reduced density matrix of the NIFG, we try
$W$ EWs corresponding to the ansatz
\begin{equation}\label{}
^{0}\mathcal{W}_{W}^{(\mathrm{sp})}=(1+\sqrt{8})\;
\textbf{I}+\widehat{P}_{12}+\widehat{P}_{13}-\widehat{P}_{23}.
\end{equation}
For the trace of the product of
$^{0}\mathcal{W}_{W}^{(\mathrm{sp})}$ with $\rho_3$ we obtain
\begin{equation}\label{}
\mathrm{Tr}(^{0}\mathcal{W}_{W}^{(\mathrm{sp})}\rho_3)=(1+\sqrt{8})-3(a+b-c),
\end{equation}
which must be negative to signal the existence of $W\setminus B$
entanglement in $\rho_3$.  For the NIFG the maximum value of
$a+b-c$ is $1$.  It can then be checked that the constraints
imposed on the coefficients $a$, $b$, and $c$ for the NIFG exclude
the possibility of a negative expectation value of
$^{0}\mathcal{W}_{W}^{(\mathrm{sp})}$ for the NIFG density matrix
operator.  We are led to conclude that symmetric spin-chain EWs
are not capable of detecting genuine tripartite entanglement in
NIFG.

On the other hand, if we consider the general case of a
density-matrix operator of the form (\ref{NIFG}) subject to the
basic constraints (\ref{constdens}), {\it but not} specializing to
the NIFG, the expectation value of
$^{0}\mathcal{W}_{W}^{(\mathrm{sp})}$ over $\rho_3$ {\it can} take
on a negative value for some set of parameter values. It should be
emphasized that entanglement detected by
$^{0}\mathcal{W}_{W}^{(\mathrm{sp})}$ in such a general density
matrix $\rho_3$ necessarily belongs to the $W\setminus B$ class
rather than the other class of genuine tripartite entanglement,
i.e., $GHZ\setminus W$. We note that in the case of the
transformed density matrix $\rho_3'$ introduced in
Eq.~(\ref{rho3prime}), the relation
\begin{align}\label{}
\mathrm{Tr}(^{0}\mathcal{W}_{W}^{(\mathrm{sp})}\rho_3')=&(1+\sqrt{8})+(-a-3b+c)|\alpha|^6\nn\\
&+(-3a-3b+3c)|\beta|^6
\nn\\&+(-8a-9b+6c)|\alpha^4\beta^2|\nn\\&+(-9a-9b+9c)|\alpha^2\beta^4|<0
\end{align}
must be satisfied for $W\setminus B$ identification.

Having explored periodic spin-chain models for EW operators, we
next describe another approach to the problem of constructing EWs
for the NIFG via LP, based on projection operators making up the
density matrix $\rho_3$ in Eq.~(\ref{NIFG}).

\subsection{$GHZ$ EWs composed of projection operators}\label{subsecB}

In this section, we develop a new class of proper $GHZ$ EWs with
the aid of projection operators (pure-state density matrices)
corresponding to the different entangled density operators arising
in tripartite systems.  If an EW is to detect tripartite
entanglement associated with a given class of biseparables (1-23,
2-13, or 3-12), or with genuine entangled states ($W\setminus B$
or $GHZ\setminus W$), it must necessarily be constructed from
operators having non-vanishing expectation value with respect to
the class that is specified. In particular, as demonstrated in the
Sec.~\ref{subsecA} (see e.g., Eq.~(\ref{spinproj})), a $GHZ$ EW
cannot be successfully built from projection operators of the
$GHZ$ set if projectors from the $GHZ\setminus W$ class are
excluded.

With this preface, we introduce the operator
\begin{align}\label{ghzEW}
&\mathcal{W}_{GHZ}^d=a_0
\textbf{I}_8\nn\\
&+a_1(|\Psi_{12}^{-}\rangle\langle\Psi_{12}^{-}|\otimes
\textbf{I}+|\Psi_{13}^{-}\rangle\langle\Psi_{13}^{-}|\otimes
\textbf{I}+\textbf{I}\otimes|\Psi_{23}^{-}\rangle\langle\Psi_{23}^{-}|
)\nn\\
&+a_2|W_1\rangle\langle W_1|+a_3|W_2\rangle\langle
W_2|+a_4|\Psi_{123}^{-}\rangle\langle\Psi_{123}^{-}|,
\end{align}
which contains projection operators for {\it all} the different
vector types involved for the three-fermion subsystem,
\begin{align}
&|W_1\rangle=\frac{1}{\sqrt{3}}(|001\rangle+|010\rangle+|100\rangle),\nonumber\\
&|W_2\rangle=\frac{1}{\sqrt{3}}(|011\rangle+|110\rangle+|101\rangle),\nonumber\\
&|\Psi_{123}^{-}\rangle=\frac{1}{\sqrt{2}}(|000\rangle-|111\rangle),
\end{align}
and the $a_i$ are real parameters. First, to ensure that the
operator (\ref{ghzEW}) qualifies as a $GHZ$ EW, we should impose
positivity of the expectation value of $\mathcal{W}_{GHZ}^d$ with
respect to $W$ states. We take the generic $W$ state vector $|
\Psi_W \rangle$ in the form (\ref{w}).  For simplicity we define
the operators
\begin{align}
&\widehat{P}_1:=2\sum_{<i,j>}|\Psi_{ij}^{-}\rangle\langle\Psi_{ij}^{-}|\otimes
\textbf{I},\nonumber\\
&\widehat{P}_2:=3|W_1\rangle\langle W_1|,\quad
\widehat{P}_3:=3|W_2\rangle\langle W_2|,\nonumber\\
&\widehat{P}_4:=2|\Psi_{123}^{-}\rangle\langle\Psi_{123}^{-}|,
\end{align}
and therewith their corresponding expectation values $P_i^{W}$ ($i=1,2,3,4$)
with respect to $W$-vectors in terms of the coefficients $A_{ijk}$
explicated in the appendix.  One readily finds that the maximum
eigenvalue is 3 for $\widehat{P}_1$, $\widehat{P}_2$, and
$\widehat{P}_3$ and 2 for $\widehat{P}_4$. A straightforward
calculation shows that $P^W_1$, $P^W_2$, and $P^W_3$ can reach their
maximum possible value, but for $P^W_4$ one finds a maximum
overlap $\langle \Psi_{123}^{-}|\psi_W\rangle$ of $3/2$.

\begin{figure}[]
\begin{center}
\includegraphics[width=7cm]{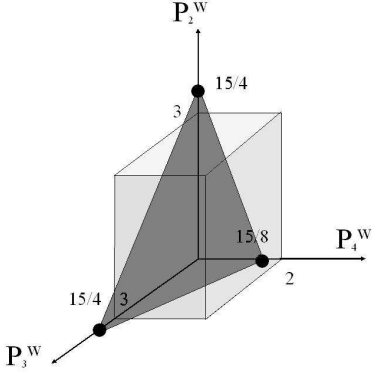}
\end{center}
\caption[]{ Projection of the feasible region of the LP problem on
the $P^W_1$ dimension, showing the intersection domain of the
region $P^W_2+P^W_3+2P^W_4\leq 15/4$ and the domain of possible
expectation values of $\widehat{P}_2, \widehat{P}_3,
\widehat{P}_4$ with respect to a generic tripartite quantum state
within the rectangular parallelepiped. } \label{PWd}
\end{figure}

To reduce the task of determining parameter values $a_i$ such that
$\mathcal{W}_{GHZ}^d$ becomes a $GHZ$ EW operator to an LP
problem, we need to find a feasible region.  Basing a first
conjecture on the maximum eigenvalues of the $\widehat{P}_i$ we
find that the extremum points
\begin{align}\label{points}
(P^W_1, P^W_2, P^W_3, P^W_4)=&(0, 0, 0, 0),(3, 0, 0, 0),(0, 3, 0,
0),\nn\\
&(0, 0, 3, 0),(0, 0, 0, 3/2)
\end{align}
cannot be vertices for a feasible region of our LP problem.  As
checked numerically, some points lying outside the convex
polyhedron with vertices specified by (\ref{points})
correspond to negative expectation values for $\mathcal{W}_{GHZ}^d$.
To compensate, we extend the proposed region by a parallel shift of
the boundary hyperplane and reduce the problem to LP as follows:
\begin{align}\label{lp2}
&\mathrm{Minimize} \quad\;\; a_0+\sum_{i=1}^4 a_i P^W_i\nonumber&\\
&\mathrm{subject\ to} \quad\;\;\left\{\begin{array}{c}
 0\leq P^W_1, P^W_2, P^W_3\leq3,\;\\ 0\leq P^W_4\leq2,\\
  0\leq P^W_1+P^W_2+P^W_3+2P^W_4\leq \frac{15}{4}. \\
\end{array}\right.&
\end{align}
Imposing the latter constraints, the operator
$\mathcal{W}_{GHZ}^d$ can still have one or more negative
expectation values since the range of the expectation value of
$\widehat{P}_1+\widehat{P}_2+\widehat{P}_3+2\widehat{P}_4$ is
bounded between 0 and 4.  Now we have a feasible region formed by
the intersection of the 4-dimensional rectangular parallelepiped
domain defined by the extremum eigenvalues of the
$\widehat{P}_i$'s and the constraining hyperplane $
P^W_1+P^W_2+P^W_3+2P^W_4\leq 15/4$. We illustrate the situation in
Fig.~\ref{PWd}, projecting onto the $P^W_1$ dimension.

The resulting convex polyhedron has vertices
\begin{widetext}
\begin{align}
(P^W_1, P^W_2, P^W_3, P^W_4)=& (0, 0, 0, 0), (3, 0, 0, 0),
(3, 0, 0, 3/8), (3, 0, 3/4, 0),\\\nn
&(3, 3/4, 0, 0), (0, 3, 0, 0), (0, 3, 0, 3/8), (0, 3, 3/4, 0),
(3/4, 3, 0, 0),\\\nn
&(0, 0, 3, 0), (0, 0, 3, 3/8), (0, 3/4, 3, 0), (3/4, 0, 3, 0),
(0, 0, 0, 15/8) .
\end{align}
\end{widetext}

After solving the LP problem posed in (\ref{lp2}), we arrive
at a set of the $a_i$,
\begin{widetext}
\begin{eqnarray}\label{ineq}
a_0\geq0, \; a_0+ 3a_1\geq0, \; a_0+ 3a_2\geq0, \; a_0+ 3a_3\geq0, \;
a_0+15a_4/8\geq0,\nonumber\\
a_0+3a_1+ 3a_4/8\geq0, \; a_0+3a_2+ 3a_4/8\geq0,
\; a_0+3a_3+ 3a_4/8\geq0, \nn\\
a_0+3a_1+3a_2/\geq0, \; a_0+3a_1+3a_3/4\geq0, \;
a_0+3a_2+3a_3/4\geq0, \nn\\
a_0+3a_2+3a_1/4\geq0, \; a_0+3a_3+3a_1/4\geq0, \;
a_0+3a_3+3a_2/4\geq0,
\end{eqnarray}
\end{widetext}
imposed by the positivity of the trace of of the EW
operator over all $W$ states.  These constraints, together with
the existence of at least one negative eigenvalue among
\begin{eqnarray}\label{eigen}
a_o, a_0 + 3a_1, a_0 + 3a_2, a_0 + 3a_3, a_0+2a_4,
\end{eqnarray}
guarantee that $\mathcal{W}_{GHZ}^d$ qualifies as a $GHZ$ EW.
Among the combinations (\ref{eigen}), only $a_0$ is excluded from
negativity, with the others remaining available for signaling a
$GHZ\setminus W$ state.

We are now prepared to search for $GHZ\setminus W$ genuine
entanglement in the NIFG. As an optimal case of our $GHZ$ EW
construction, the inequality in Eq.~(\ref{lp2}) specifying the
last boundary hyperplane of the present LP problem yields the
explicit $GHZ$ EW operator
\begin{align}\label{}
^{0}\mathcal{W}_{GHZ}^d=&\frac{15}{4}
\textbf{I}_8-2\sum_{<i,j>}|\Psi_{ij}^{-}\rangle\langle\Psi_{ij}^{-}|-3|W_1\rangle\langle
W_1|\nn\\&-3|W_2\rangle\langle
W_2|-4|\Psi_{123}^{-}\rangle\langle\Psi_{123}^{-}|.
\end{align}
Adopting the witness $^{0}\mathcal{W}_{GHZ}^d$ for our search,
we calculate its expectation value for the three-fermion reduced
density matrix $\rho_3$ of the NIFG, obtaining the simple
expression
\begin{equation}\label{w0}
\mathrm{Tr}(^{0}\mathcal{W}_{GHZ}^d \rho_3)=\frac{1}{2}+2\eta,
\end{equation}
which would have to be negative to confirm the presence of
$GHZ\setminus W$ entanglement.  However, as seen in Fig.~\ref{Tr},
the term $\eta$ does not reach negative values, ruling out this
possibility. Moreover, calculation of the expectation value of the
best-case witness operator $\mathcal{W}_{GHZ}^d$ with respect to
the rotated three-fermion reduced density matrix of
Eq.~(\ref{rho3prime}) yields
\begin{align}\label{trdens}
&\mathrm{Tr}(^{0}\mathcal{W}_{GHZ}^d
\rho_3')=\frac{15}{4}\nn\\
&+\left(\frac{|\alpha^4
\beta^2|}{4}+\frac{|\alpha^2 \beta^4|}{4}\right)\left(-33+9a-3b+9c\right)\nn\\
&-\left(\frac{|\alpha|^6}{4}+\frac{|\beta|^6}{4}\right)
\left(11+a+b+c\right)\nn\\
&+\left(\frac{\alpha^{*3} \beta^3}{2}+\frac{\beta^{*3}
\alpha^3}{2}\right) (1-a-b-c),
\end{align}
which again fails to attain negative values.

\begin{figure}[]
\begin{center}
\includegraphics[width=7.8cm]{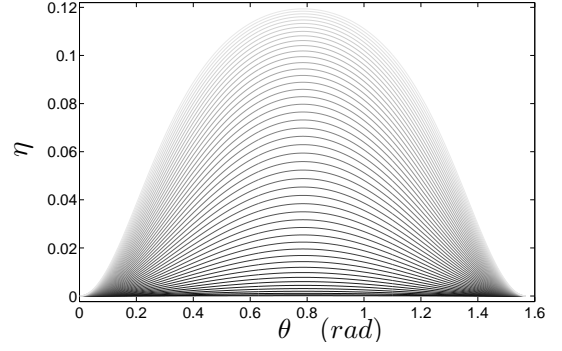}
\end{center}
\caption[]{Plots of the term $\eta$ in Eq.~(\ref{w0}) versus
$\theta$ (in radians) at different values of $k_Fr$ for the
2-d configuration shown in Fig.~\ref{p_2d}. (Plot traces change
from black to light gray as $k_F$ is increased from 0.1 to 5.0
in steps of 0.1.)
} \label{Tr}
\end{figure}

In the next subsection, $GHZ$ stabilizer operators will be
employed to formulate still another class of entanglement witnesses.

\subsection{Stabilizer EWs}
Previous work has demonstrated the utility and robustness of EWs
based on stabilization operators \cite{Gottes,Toth1,Toth2,hes2}.
By definition a stabilizer operator ${\hat S}$ for state $ |\psi
\rangle$ has the property ${\hat S} | \psi \rangle = | \psi
\rangle$. T\'oth and G\"uhne \cite{Toth1,Toth2} have shown that if
some of the stabilizing operators for a given state are available,
entanglement conditions may be found that detect states in the
neighborhood of this state.  Here we apply the stabilizer
formalism to obtain a parameterized EW for detecting quantum
states close to a $GHZ$ entangled state. Also employing stabilizer
operators, we are able to discriminate between different types of
genuine tripartite entanglement that could be present in the
three-particle reduced density matrix.

We begin by considering a linear combination
\begin{equation}\label{stab}
\mathcal{W}^{({\rm stab})}=b_0 \textbf{I}+b_1 \widehat{S}_1+b_2
\widehat{S}_2+b_{1,2}\widehat{S}_{1,2}
\end{equation}
of $GHZ$ stabilizer operators
$\widehat{S}_1=\sigma^{(1)}_x\sigma^{(2)}_x\sigma^{(3)}_x$,
$\widehat{S}_2=\sigma^{(1)}_z\sigma^{(2)}_z\textbf{I}^{(3)}$, and
$\widehat{S}_{1,2}=\widehat{S}_1\times\widehat{S}_2$, where the
$b_i$'s are real parameters.  First, to find ranges of the
parameters $b_i$ such that $\mathcal{W}^{({\rm stab})}$ becomes a
$W$ EW, we must determine the domain spanned by the expectation
values $S^B_i$ of the corresponding operators $\widehat{S}_i$ over
the biseparable set of states.  Straightforward calculation shows
that this task can be reduced to the solution of the following LP
problem (see the appendix):
\begin{widetext}
\begin{align}\label{boundstab}
&\mathrm{Minimize} \quad\;\; b_0+b_{1} S^B_{1}+b_{2} S^B_{2}+b_{12} S^B_{12}\nonumber\\
&\mathrm{subject\ to} \quad\;\;\left\{\begin{array}{cc}
\begin{array}{c}
(-1)^{i_1}S^B_1+(-1)^{i_2}S^B_2+(-1)^{i_1+i_2}S^B_{1,2}\leq\sqrt{2},\\
(-1)^{i_1}S^B_1+(-1)^{i_2}S^B_2+(-1)^{i_1+i_2+1}S^B_{1,2}\leq1,
\end{array} &
\;\;\;\;\forall(i_1,i_2)\in\{0,1\}^2.
\end{array}\right.
\end{align}
\end{widetext}
It is to be noted here that the boundaries associated with
separable states cannot exceed unity \cite{hes2}.
The operators corresponding to the second set
of constraints in (\ref{boundstab}), i.e.,
\begin{equation}\label{posops}
(-1)^{i_1}\widehat{S}_1+(-1)^{i_2}\widehat{S}_2
+(-1)^{i_1+i_2+1}\widehat{S}_{1,2},\quad (i_1,i_2)\in\{0,1\},
\end{equation}
are positive operators and thus cannot form EW operators.
In solving the LP problem stated in (\ref{boundstab}), we find
that the constraints
\begin{widetext}
\begin{equation}\label{Wconstraint}
\begin{array}{cc}
b_0+\sqrt{2}(b_1+b_2-b_{1,2})\geq0, &
b_0+\sqrt{2}(b_1-b_2+b_{1,2})\geq0,\\
b_0+\sqrt{2}(-b_1+b_2+b_{1,2})\geq0, & b_0-\sqrt{2}(b_1+b_2+b_{1,2})\geq0, \\
b_0+b_1-b_2-b_{1,2}\geq0, & b_0-b_1-b_2+b_{1,2}\geq0,\\
b_0-b_1+b_2-b_{1,2}\geq0, & b_0+b_1+b_2+b_{1,2}\geq0\\
b_0+\sqrt{2}b_1+\frac{1-\sqrt{2}}{2}b_2+\frac{-1+\sqrt{2}}{2}b_{1,2}\geq0, & b_0+\sqrt{2}b_1+\frac{-1+\sqrt{2}}
{2}b_2+\frac{1-\sqrt{2}}{2}b_{1,2}\geq0,\\
b_0+\frac{-1+\sqrt{2}}{2}b_1+\sqrt{2}b_2+\frac{1-\sqrt{2}}{2}b_{1,2}\geq0, & b_0+\frac{1-\sqrt{2}}{2}b_1+\sqrt{2}
b_2+\frac{-1+\sqrt{2}}{2}b_{1,2}\geq0,\\
b_0+\frac{-1+\sqrt{2}}{2}b_1+\frac{1-\sqrt{2}}{2}b_2+\sqrt{2}b_{1,2}\geq0, & b_0+\frac{-1+\sqrt{2}}{2}b_1+
\frac{1-\sqrt{2}}{2}b_2+\sqrt{2}b_{1,2}\geq0,\\
b_0+\frac{1-\sqrt{2}}{2}b_1+\frac{1-\sqrt{2}}{2}b_2-\sqrt{2}b_{1,2}\geq0, & b_0+\frac{-1+\sqrt{2}}{2}b_1+
\frac{-1+\sqrt{2}}{2}b_2-\sqrt{2}b_{1,2}\geq0,\\
b_0+\frac{1-\sqrt{2}}{2}b_1-\sqrt{2}b_2+\frac{1-\sqrt{2}}{2}b_{1,2}\geq0, & b_0+\frac{-1+\sqrt{2}}{2}b_1-
\sqrt{2}b_2+\frac{-1+\sqrt{2}}{2}b_{1,2}\geq0,\\
b_0-\sqrt{2}b_1+\frac{1-\sqrt{2}}{2}b_2+\frac{1-\sqrt{2}}{2}b_{1,2}\geq0,
&
b_0-\sqrt{2}b_1+\frac{-1+\sqrt{2}}{2}b_2+\frac{-1+\sqrt{2}}{2}b_{1,2}\geq0\\
\end{array}
\end{equation}
\end{widetext}
guarantee positivity of $\mathcal{W}^{({\rm stab})}$ over all
biseparable states. However, for $\mathcal{W}^{({\rm stab})}$ to
qualify as the $W$ EW we seek, it also must possess at least one
negative eigenvalue from the possibilities
\begin{equation}\label{eigenstab}
b_0+(-1)^{i_1}b_1+(-1)^{i_2}b_2+(-1)^{i_1+i_2}b_{1,2}, \quad
\forall(i_1,i_2)\in\{0,1\}^2
\end{equation}

Next, to enable discrimination between different kinds of genuine
entangled states, a $GHZ$ EW is required.  Accordingly, we should
find values of the $b$ coefficients in Eq.~(\ref{stab}) such that
$\mathcal{W}^{({\rm stab})}$ is positive over all the states in
the $W$ class and yet has at least one negative eigenvalue.  To
this end, we search for a polyhedron spanned by the expectation
values $S^W_1$, $S^W_2$, and $S^W_{1,2}$, which are functions of
the coefficients $A_{ijk}$ in the general $W$ vector
$|\Psi_W\rangle$ as expressed previously. Following the same
pattern as before, we are led to the LP problem
\begin{widetext}
\begin{align}\label{LPstabW}
&\mathrm{Minimize} \quad\;\; b_0+b_{1} S^W_{1}+b_{2} S^W_{2}+b_{12} S^W_{12}\nonumber\\
&\mathrm{subject\ to} \quad\;\;\left\{\begin{array}{cc}
\begin{array}{c}
(-1)^{i_1}S^W_1+(-1)^{i_2}S^W_2+(-1)^{i_1+i_2}S^W_{1,2}\leq 2.98\\
(-1)^{i_1}S^W_1+(-1)^{i_2}S^W_2+(-1)^{i_1+i_2+1}S^W_{1,2}\leq1
\end{array} &
\;\;;\;\;\forall(i_1,i_2)\in\{0,1\}^2.
\end{array}\right.
\end{align}
\end{widetext}
In this case, operators corresponding to the first cluster of the
constraints in (\ref{LPstabW}), i.e.,
\begin{equation}
(-1)^{i_1}\widehat{S}_1+(-1)^{i_2}\widehat{S}_2+(-1)^{i_1+i_2}\widehat{S}_{1,2}
\end{equation}
can serve as $GHZ$ EW operators. Solution of the LP problem
(\ref{LPstabW}) yields a new set of constraints on the $b$
parameters, namely
\begin{widetext}
\begin{equation}\label{GHZconstraint}
\begin{array}{cc}
b_0+2.98(b_1+b_2-b_{1,2})\geq0, & b_0+2.98(b_1-b_2+b_{1,2})\geq0,\\
b_0+2.98(-b_1+b_2+b_{1,2})\geq0, & b_0-2.98(b_1+b_2+b_{1,2})\geq0, \\
b_0+b_1-b_2-b_{1,2}\geq0, & b_0-b_1-b_2+b_{1,2}\geq0, \\
b_0-b_1+b_2-b_{1,2}\geq0, & b_0+b_1+b_2+b_{1,2}\geq0\\
b_0+2.98 b_1-0.99 b_2+0.99 b_{1,2}\geq0, & b_0+2.98 b_1+0.99 b_2-0.99 b_{1,2}\geq0,\\
b_0+0.99 b_1+2.98 b_2-0.99 b_{1,2}\geq0, & b_0-0.99 b_1+2.98 b_2+0.99 b_{1,2}\geq0,\\
b_0+0.99 b_1-0.99 b_2+2.98 b_{1,2}\geq0, & b_0-0.99 b_1+0.99 b_2+2.98 b_{1,2}\geq0,\\
b_0-0.99 b_1-0.99 b_2-2.98 b_{1,2}\geq0, & b_0+0.99 b_1+0.99 b_2-2.98 b_{1,2}\geq0,\\
b_0-0.99 b_1-2.98 b_2-0.99 b_{1,2}\geq0, & b_0+0.99 b_1-2.98 b_2+0.99 b_{1,2}\geq0,\\
b_0-2.98 b_1-0.99 b_2-0.99 b_{1,2}\geq0, & b_0-2.98b_1+0.99
b_2+0.99 b_{1,2}\geq0,
\end{array}
\end{equation}
\end{widetext}
required for $\mathcal{W}^{({\rm stab})}$ to have only positive
expectation values over $W$ class.

To construct an EW operator $\mathcal{W}^{({\rm stab})}$ which can
detect a genuine entangled state belonging to $W\setminus B$ but
not accept a state from $GHZ\setminus W$, the parametric
constraints (\ref{Wconstraint}) should be satisfied, while some
constraint among the set (\ref{GHZconstraint}) should be violated.
To fulfill the requirement for detection, at least one of the
eigenvalues in the set (\ref{eigenstab}) must be negative.

Among the above EWs of the form (\ref{stab}), we try the following
\begin{align}\label{0Wstab}
^{0}\mathcal{W}^{({\rm stab})}_W=\sqrt{2}\;\textbf{I}+\widehat{S}_1+\widehat{S}_2-\widehat{S}_{12},\nn\\
^{0}\mathcal{W}^{({\rm stab})}_{GHZ}=2.98\;\textbf{I}+\widehat{S}_1+\widehat{S}_2-\widehat{S}_{12}.
\end{align}
To test for a genuine tripartite entanglement in the NIFG, we
evaluate the expectation value of the first of these operators
with respect to the reduced density matrix $\rho_3$. For
$GHZ\setminus B$ entanglement to be present, the quantity
\begin{equation}\label{}
\mathrm{Tr}(^{0}\mathcal{W}^{({\rm stab})}_W\rho_3)=\sqrt{2}-2a.
\end{equation}
must be negative, i.e.,  $a>1/\sqrt{2}$.  Referring to
Fig.~\ref{p_2d}, this would occur for negative $b$ in the 2-d
configuration.

To detect $GHZ\setminus W$ entanglement in the NIFG, the quantity
\begin{equation}\label{}
\mathrm{Tr}(^{0}\mathcal{W}^{({\rm stab})}_{GHZ}\rho_3)=2.98-2a
\end{equation}
should reach negative values, {\it but this is not possible since}
$|a|<1$. Considering the EWs defined in Eq.~(\ref{0Wstab}), we see
that for a $W\setminus B$ entangled state to be detected, the
expectation value of
$\widehat{S}_1+\widehat{S}_2-\widehat{S}_{12}$ must lie between
$\sqrt{2}$ and $2.98$; for the NIFG this requires that the
condition $-1.49<a<-\sqrt{2}/2$ is fulfilled. Importantly, upon
referring to Fig.~\ref{Tr}, we confirm that $GHZ\setminus W$
entanglement does not exist in the three-fermion density matrix
$\rho_3$ of the NIFG, and that all the states
detectable in $\rho_3$ by the entanglement witness
$^{0}\mathcal{W}^{({\rm stab})}_{GHZ}$ belong to the $GHZ\setminus
W$ set.

\begin{figure}[]
\begin{center}
\includegraphics[width=8.5cm]{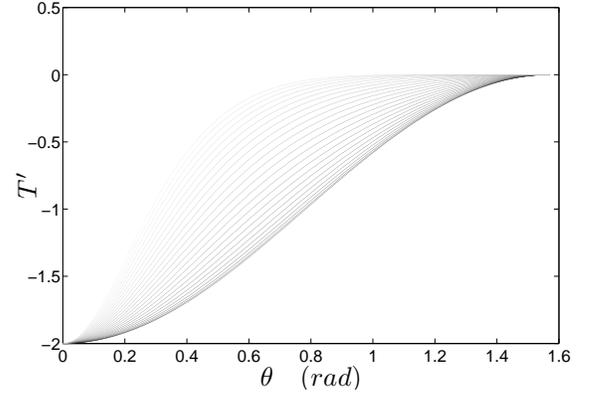}
\end{center}
\caption[]{Plot of the trace
$T'=\mathrm{Tr}\{(\widehat{S}_{1}+\widehat{S}_{2}-\widehat{S}_{12})\rho_3\}$
versus $\theta$ (in radians) at different values of $k_Fr$ for the
2-d configuration shown in Fig.~\ref{p_2d}. (Plot traces change
from black to light gray as $k_Fr$ is increased from 0.2 to 5.0 in
steps of 0.2.) } \label{Tr}
\end{figure}

\begin{figure}[]
\centering
\subfigure[]{
\includegraphics[width=8cm]{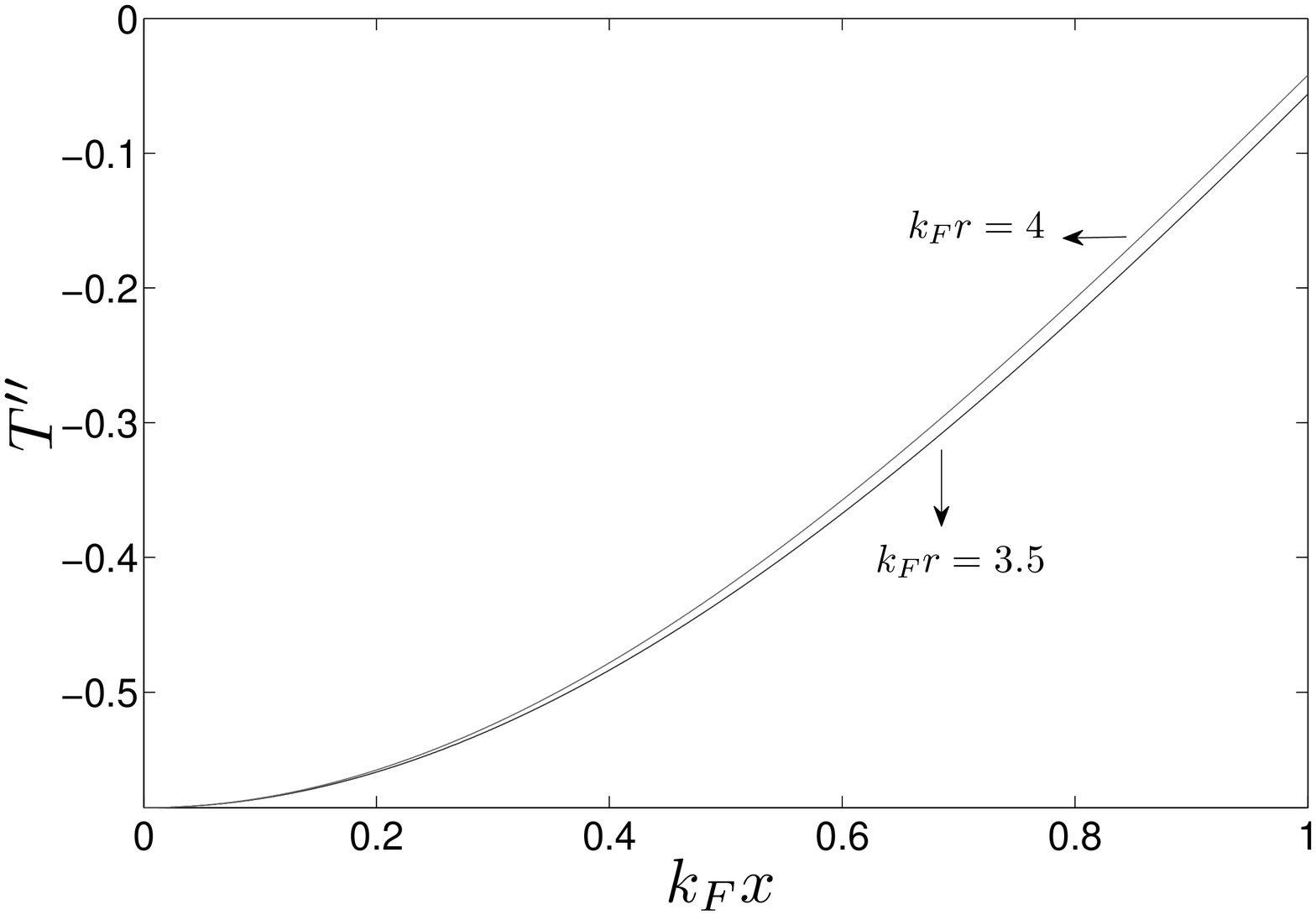}}
\subfigure[]{
\includegraphics[width=8cm]{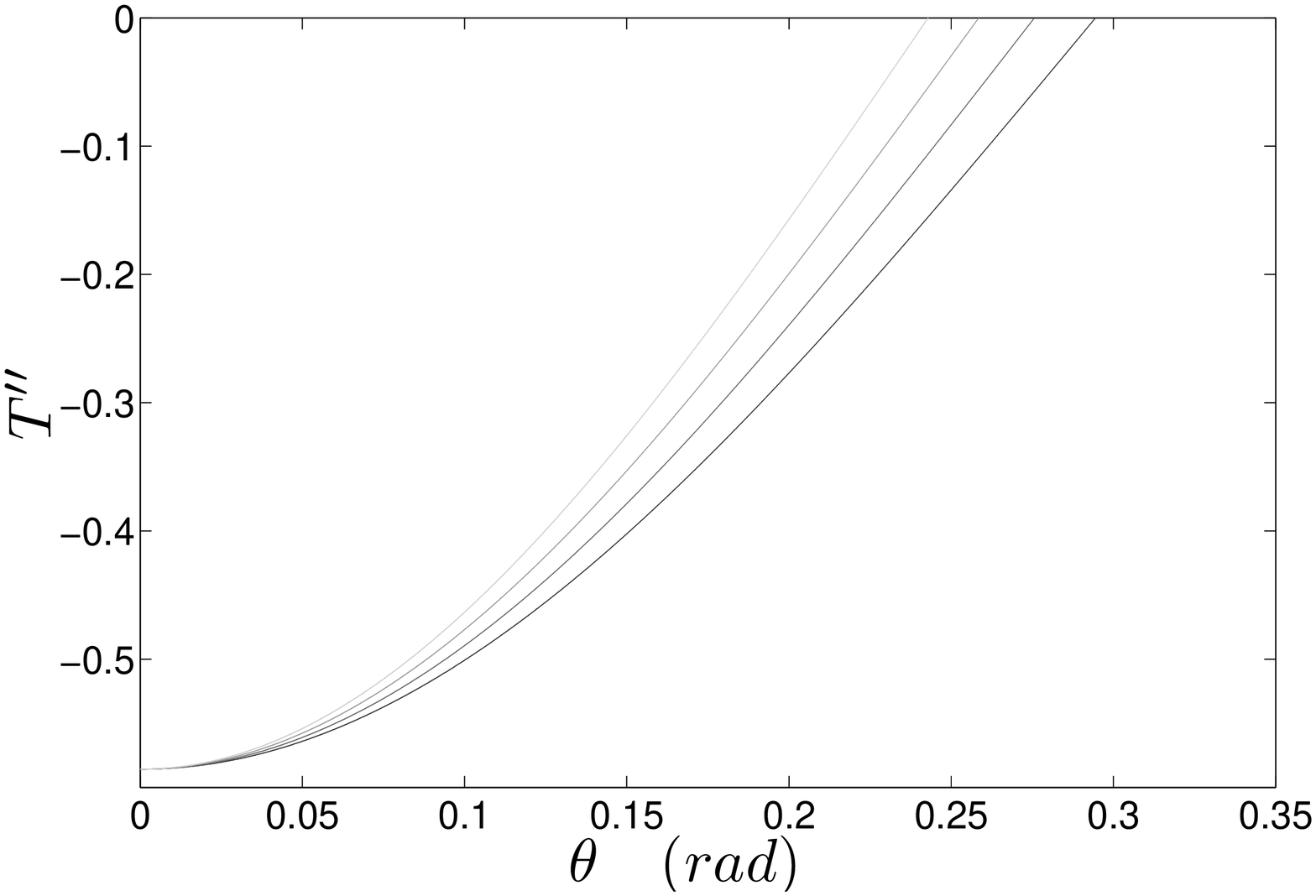}}
\caption[]{ Trace $T''=\mathrm{Tr}\{^{0}\mathcal{W}^{({\rm
stab})}_{W}\rho_3\}$ for detection of PPT genuine tripartite
$W\setminus B$ entanglement with respect to the third particle.
(a) The trace $T''$ is plotted versus $k_Fx$ for the 1-d
configuration shown in Fig.~\ref{p_1d} at $k_Fr=3.0$ and
$k_Fr=3.5$. (b) Trace $T''$ is plotted versus $\theta$ for the 2-d
configuration shown in Fig.~\ref{p_2d} for $3.0\leq k_Fr\leq4.0$
in steps of $0.25$. (Plot traces change from black to light gray
as $k_Fr$ is increased from 3.0 to 4.0.)} \label{PPT}
\end{figure}

A measure called negativity has been employed in Ref.~\cite{Lunk}
in studying the entanglement properties of the NIFG. This quantity
is defined for a trio of fermions $i,j,k$ as $N_{[i,jk]} =
(||\rho_3^{T_i}||_1 -1)/2$, where $||\rho_3^{T_i}||_1$ is the
trace norm of the partial transpose of the three-fermion reduced
density matrix $\rho_3$ of fermion $i$ with respect to fermions
$j,k$. One of the advantages of working with EW operators rather
than the negativity measure is in the identification of PPT
entangled states. Using the condition (\ref{cond3}) on the
coefficients $a$, $b$, and $c$ in the NIFG expression for $\rho_3$
together with the entanglement witness $^{0}\mathcal{W}^{({\rm
stab})}_{W}$, one can determine the configuration domain for which
there exists PPT genuine $W$ entanglement with respect to the
third fermion. The behavior of the trace
$T''=\mathrm{Tr}\{^{0}\mathcal{W}^{({\rm stab})}_{W}\rho_3\}$ at
three values of $k_Fr$ is shown in Fig.~\ref{PPT} for both 1-d and
2-d configurations. Hence $^{0}\mathcal{W}^{({\rm stab})}_{W}$ is
a partially non-decomposable $W$ EW for the third party since it
can detect a PPT state with respect to the third fermion.

It is instructive to note that the eigenvectors corresponding to
the eigenvalues of $\rho_3$ in (\ref{NIFG}) belong to the
$W\setminus B$ class rather than $GHZ\setminus W$, affirming the
absence of genuine $GHZ\setminus W$ entanglement in the NIFG. By
way of proof, we first rewrite the three-body reduced density
matrix in the standard form $\rho_3=\sum_i d_i
|\alpha_i\rangle\langle \alpha_i|$, in which the $d_i$ are
necessarily nonnegative.  Then we assume there exists a $GHZ$ EW
operator $\mathcal{W}_{GHZ}$ such that
\begin{equation}\label{}
\mathrm{Tr}(\mathcal{W}_{GHZ}\rho_3)=\sum_i d_i
\mathrm{Tr}(\mathcal{W}_{GHZ}|\alpha_i\rangle\langle \alpha_i|)
\end{equation}
is negative. This assumption is contradictory if there is no
$GHZ\setminus W$ contribution to the eigenvectors of $\rho_3$.

One question still remains (cf.~Ref.~\onlinecite{Ved1}): can there
exist tripartite $W$-type entanglement in the NIFG, without any
admixture biseparable or separable components, i.e., a state
belonging purely to the $W\setminus B$ subset?  This aspect can be
investigated by establishing an upper bound on the trace of the
product of a general density operator $\rho^{W}$ possessing
$W\setminus B$ entanglement, and the three-fermion reduced density
operator $\rho_3$.  By virtue of Fermi exchange antisymmetry, it
is sufficient to work with the operator
\begin{align}\label{}
\varrho^W=&|W_1\rangle\langle
W_1|\nn\\=&\frac{1}{3}(|001\rangle+|010\rangle+|100\rangle)(\langle
001|+\langle 010|+\langle 100|),
\end{align}
together with its transform
\begin{equation}\label{}
\varrho'^{W}=\left(
  \begin{array}{cc}
    \beta^* & \alpha \\
    -\alpha^* & \beta \\
  \end{array}
\right)^{\otimes 3}\varrho^{W}\left(
  \begin{array}{cc}
    \beta & -\alpha \\
    \alpha^* & \beta^* \\
  \end{array}
\right)^{\otimes 3}
\end{equation}
under an arbitrary local unitary transformation, which again
yields a purely $W$-type genuine entangled state.
We have verified numerically that the value of $\mathrm{Tr}(\varrho'^{W}
\rho_3)$ is always less than one; hence the three-fermion reduced
density operator of the NIFG cannot take the ``pure''
form $\varrho'^W$.  This finding is confirmed for the explicit forms
of $\rho(s,s',s'';t,t',t'')$ and $\rho_3$ given in
Ref.~\onlinecite{Ved1} and reproduced in Eqs.~(\ref{fff}) and
(\ref{rho3matrix}).  Consequently, one can conclude that in the
NIFG, genuine tripartite entanglement only occurs in the three-fermion
reduced density matrix $\rho_3$ in the company of 1-separable
and/or 2-separable entanglement in other partitions: one cannot
generate ``pure'' tripartite entanglement in the noninteracting Fermi gas.

\section{Conclusion}
We have introduced new classes of entanglement witnesses (EWs) for
the purpose of identifying genuine tripartite entanglement, i.e.,
$W\setminus B$ and $GHZ\setminus W$ states, in the three-fermion
density matrix of the noninteracting Fermi gas (NIFG).  We have
reduced the task of constructing suitable EWs for this system to
well-defined problems of linear programming. Considering EW
operators inspired by a periodic spin chain model, EWs composed of
the projection operators over the different classes of tripartite
systems, and $GHZ$ stabilizer operators, we have found that the
genuine tripartite entanglement present in the NIFG belongs to the
$W\setminus B$ class. This result is confirmed in the structure of
the eigenvectors of a general three-fermion reduced density matrix
$\rho_3$ of the NIFG. We have seen that genuine tripartite
entanglement does not occur in ``pure'' form, but instead it
appears mixed with $B$ or $S$ components. Additionally, using a
partially non-decomposable EW, we have been able to detect PPT
genuine $W\setminus B$ entanglement with respect to the third
party of a fermion trio in the NIFG. The general approach followed
in this work can be applied to investigate multipartite
entanglement in other quantum many-particle systems, whether
consisting of fermions or bosons, possibly with higher spins, and
whether the particles are interacting or noninteracting.

\begin{center}
\textbf{Acknowledgements}
\end{center}
The authors thank D. Bru{\ss} and T. V\'{e}rtesi for useful
comments and M. A. Jafarizadeh, G. Najarbashi, and B. Dastmalchi
for motivation and informative discussions. This research has been
supported by the European Community Project N2T2. J.W.C. is
grateful to the Johannes Kepler Universit\"at Linz for fellowship
support and the Institut f\"ur Theoretical Physik for hospitality
during a sabbatical leave.  He also thanks the Complexo
Interdisciplinar of the University of Lisbon and the Department of
Physics of the Technical University of Lisbon for their gracious
hospitality, while acknowledging research support from
Funda\c{c}\~{a}o para a Ci\^{e}ncia e a Tecnologia of the
Portuguese Minist\'erio da Ci\^{e}ncia, Tecnologia e Ensino
Superior as well as Funda\c{c}\~{a}o Luso-Americana.

\renewcommand{\theequation}{A-\arabic{equation}}
\setcounter{equation}{0}  

\begin{center}
\section*{Appendix}
\end{center}

To obtain the most general form for quantum states in the $W$ set,
we consider the explicit form of $|\psi_W\rangle$ given
in Eq.~(\ref{w}), together with an arbitrary local $SU(2)$ transformation
applied for the different parties $i$ according to
$|0\rangle_i\rightarrow \alpha_i|0\rangle_i+\beta_i|1\rangle_i$
and $|1\rangle_i\rightarrow \beta_i^*|0\rangle_i-\alpha_i^*|1\rangle_i$,
with $|\alpha_i|^2+|\beta_i|^2=1$.  We next write the general
locally transformed $W$ vector as $|\Psi_W\rangle:=V|\psi_W\rangle$,
where
\begin{equation}\label{}
V=\left(
\begin{array}{cc}
    \beta_1^* & \alpha_1 \\
    -\alpha_1^* & \beta_1 \\
  \end{array}
\right)\otimes\left(
\begin{array}{cc}
    \beta_2^* & \alpha_2 \\
    -\alpha_2^* & \beta_2 \\
  \end{array}
\right)\otimes\left(
\begin{array}{cc}
    \beta_3^* & \alpha_3 \\
    -\alpha_3^* & \beta_3 \\
  \end{array}
\right).
\end{equation}
Rewriting the general $W$ vector as an expansion
$|\Psi_W\rangle=\sum_{i,j,k=0,1}A_{ijk}|ijk\rangle$,
the coefficients $A_{ijk}$ are determined as
\begin{align}
&A_{000}=\lambda_0\alpha_1\alpha_2\alpha_3+\lambda_1\beta_1^*\alpha_2 \alpha_3+\lambda_2\beta_1^*\alpha_2\beta_3^*+\lambda_3\beta_1^*\beta_2^*\alpha_3,  \nonumber\\
&A_{001}=\lambda_0\alpha_1\alpha_2\beta_3+\lambda_1 \beta_1^*\alpha_2\beta_3-\lambda_2\beta_1^*\alpha_2\alpha_3^*+\lambda_3\beta_1^*\beta_2^*\beta_3,  \nonumber\\
&A_{010}=\lambda_0\alpha_1\beta_2\alpha_3+\lambda_1\beta_1^*\beta_2 \alpha_3+\lambda_2\beta_1^*\beta_2\beta_3^*-\lambda_3\beta_1^*\alpha_2^*\alpha_3,\nonumber\\
&A_{011}=\lambda_0\alpha_1\beta_2\beta_3+\lambda_1\beta_1^*\beta_2\beta_3-\lambda_2\beta_1^*\beta_2\alpha_3^*-\lambda_3\beta_1^*\alpha_2^*\beta_3 , \nonumber\\
&A_{100}=\lambda_0\beta_1\alpha_2\alpha_3-\lambda_1\alpha_1^*\alpha_2 \alpha_3-\lambda_2\alpha_1^*\alpha_2\beta_3^*-\lambda_3\alpha_1^*\beta_2^*\alpha_3,  \nonumber\\
&A_{101}=\lambda_0\beta_1\alpha_2\beta_3-\lambda_1\alpha_1^*\alpha_2 \beta_3+\lambda_2\alpha_1^*\alpha_2\alpha_3^*-\lambda_3\alpha_1^*\beta_2^*\beta_3,  \nonumber\\
&A_{110}=\lambda_0\beta_1\beta_2\alpha_3-\lambda_1\alpha_1^*\beta_2 \alpha_3-\lambda_2\alpha_1^*\beta_2\beta_3^*+\lambda_3\alpha_1^*\alpha_2^*\alpha_3,  \nonumber\\
&A_{111}=\lambda_0\beta_1\beta_2\beta_3-\lambda_1\alpha_1^*\beta_2\beta_3+\lambda_2\alpha_1^*\beta_2\alpha_3^*+\lambda_3\alpha_1^*\alpha_2^*\beta_3.  \nonumber\\
\end{align}

For the expectation values of the operators $\widehat{P}_{ij}$ of
Eq.~(\ref{Pij}) over the $W$ set, we now have
\begin{align}\label{}
P^W_{12}:=&\langle\Psi_W|\widehat{P}_{12}|\Psi_W\rangle
\nn\\=&|A_{000}|^2+|A_{001}|^2-|A_{010}|^2-|A_{011}|^2\nn\\
&-|A_{100}|^2-|A_{101}|^2+|A_{110}|^2+|A_{111}|^2\nn\\
&+(A_{010}A_{100}^{\ast}+A_{011}A_{101}^{\ast}+{\rm c.c}),\nn\\
P^W_{23}:=&\langle\Psi_W|\widehat{P}_{23}|\Psi_W\rangle
\nn\\=&|A_{000}|^2-|A_{001}|^2-|A_{010}|^2+|A_{011}|^2\nn\\
&+|A_{100}|^2-|A_{101}|^2-|A_{110}|^2+|A_{111}|^2\nn\\
&+(A_{001}A_{010}^{\ast}+A_{101}A_{110}^{\ast}+{\rm c.c.}),\nn\\
P^W_{13}:=&\langle\Psi_W|\widehat{P}_{13}|\Psi_W\rangle
\nn\\=&|A_{000}|^2-|A_{001}|^2+|A_{010}|^2-|A_{011}|^2\nn\\
&-|A_{100}|^2+|A_{101}|^2-|A_{110}|^2+|A_{111}|^2\nn\\
&+(A_{000}A_{101}^{\ast}+A_{001}A_{100}^{\ast}+A_{010}A_{111}^{\ast}+A_{011}A_{110}^{\ast}
\nn\\&+{\rm c.c.}),
\end{align}
while for the $W$-set expectation values of the stabilization operators
$\widehat{S}_i$'s appearing in Eq.~(\ref{stab}) we obtain
\begin{align}\label{}
S^W_1:=&(A_{000}A_{111}^{\ast}+A_{001}A_{110}^{\ast}+A_{010}A_{101}^{\ast}\nn\\
&+A_{011}A_{100}^{\ast}+{\rm c.c.}),\nn\\
S^W_2:=&|A_{000}|^2+|A_{001}|^2-|A_{010}|^2-|A_{011}|^2\nn\\
&-|A_{100}|^2-|A_{101}|^2+|A_{110}|^2+|A_{111}|^2,\nn\\
S^W_{1,2}:=&(-A_{000}A_{110}^{\ast}-A_{001}A_{111}^{\ast}+A_{010}A_{100}^{\ast}\nn\\
&+A{011}A_{101}^{\ast}+A_{100}A_{010}^{\ast}
+A_{101}A_{011}^{\ast}\nn\\
&-A_{110}A_{000}^{\ast}-A_{111}A_{001}^{\ast}+ {\rm c.c.}).
\end{align}
Furthermore, for the operators $\widehat{P}_{i}$'s entering
Eq.~(\ref{witdef}), the expectation values over the states of the $W$ class
read
\begin{align}
P^W_1=&2\langle
\Psi_W|(|\Psi_{12}^{-}\rangle\langle\Psi_{12}^{-}|\otimes
\textbf{I}+|\Psi_{13}^{-}\rangle\langle\Psi_{13}^{-}|\otimes
\textbf{I}\nn\\
&+\textbf{I}\otimes|\Psi_{23}^{-}\rangle\langle\Psi_{23}^{-}|
)|\Psi_W \rangle\nn\\
=&(A_{001}^*+A_{010}^*+A_{100}^*)
\times(A_{001}+A_{010}+A_{100}),\nonumber\\
P^W_2=&3|\langle \Psi_W|
W_1\rangle|^2\nn\\
=&(A_{001}^*+A_{010}^*+A_{100}^*)
\times(A_{001}+A_{010}+A_{100}),\nonumber\\
P^W_3=&3|\langle \Psi_W|
W_2\rangle|^2\nn\\
=&(A_{011}^*+A_{110}^*+A_{101}^*)
\times(A_{011}+A_{110}+A_{101}),\nonumber\\
P^W_4=&2|\langle
\Psi_{123}^{-}|\Psi_W\rangle|^2\nn\\=&(A_{000}^*-A_{111}^*)\times(A_{000}-A_{111}).
\end{align}

Similar relations are generated for the expectation values of the
operators $P_{ij}^B$, $S^B_i$, and $P^B_{i}$.

To find a domain of the parameters in Eq.~(\ref{Wspin1}) such that
$\mathcal{W}^{({\rm sp})}$ qualifies as a $W$ EW, we search for a
convex polyhedron which is embedded in the domain spanned by the
$P^B_{ij}$. To illustrate how the problem of finding a $W$ EW
based on $\mathcal{W}^{(sp)}$ is reduced to the LP problem stated
in (\ref{lp3}), one of the boundaries specified in Eq.~(\ref{lp3})
for the convex polyhedron is determined as follows, the pattern
for the others being similar \cite{Guh1}.  With
$|\psi_B\rangle=|\psi_1\rangle|\psi_{23}\rangle$ denoting an
arbitrary biseparable state having entanglement among the second
and third parties, we have
\begin{align}\label{}
&|-P^B_{12}-P^B_{13}+P^B_{23}|= \nn\\
&|\langle\psi_B|\sigma^{(1)}_x\sigma^{(2)}_x+\sigma^{(1)}_y\sigma^{(2)}_y+\sigma^{(1)}_z\sigma^{(2)}_z
+\sigma^{(3)}_x+\sigma^{(3)}_y+\sigma^{(3)}_z\nn\\
&+\sigma^{(2)}_x\sigma^{(3)}_x+\sigma^{(2)}_y\sigma^{(3)}_y+\sigma^{(2)}_z\sigma^{(3)}_z
|\psi_B\rangle| \nn\\
&\leq\;|\langle\psi_B|\sigma^{(2)}_x+\sigma^{(2)}_y+\sigma^{(2)}_z+\sigma^{(3)}_x+\sigma^{(3)}_y+\sigma^{(3)}_z
|\psi_B\rangle|\nn\\
&+|\langle\psi_B|\sigma^{(2)}_x\sigma^{(3)}_x+\sigma^{(2)}_y\sigma^{(3)}_y+\sigma^{(2)}_z\sigma^{(3)}_z
|\psi_B\rangle|\nn\\&\leq \; 1+\sqrt{8}.
\end{align}
The Schwartz inequality has been invoked in the second step.
Parallel considerations apply in developing the other EWs introduced
in this paper.
In particular, for one of the boundaries associated with the stabilizer
EW corresponding to Eq.~(\ref{boundstab}) we obtain
\begin{align}\label{}
&|-S^B_{1}+S^B_{2}-S^B_{12}|=\nn\\
&\;|\langle\psi_B|-\sigma^{(1)}_x\sigma^{(2)}_x\sigma^{(3)}_x
+\sigma^{(1)}_z\sigma^{(2)}_z+\sigma^{(1)}_y\sigma^{(2)}_y\sigma^{(3)}_x|\psi_B\rangle|\nn\\
&\leq\; |\langle\psi_B|\sigma^{(2)}_x\sigma^{(3)}_x|\psi_B\rangle|
+|\langle\psi_B|\sigma^{(2)}_z|\psi_B\rangle|\nn\\
&+|\langle\psi_B|\sigma^{(2)}_y\sigma^{(3)}_x|\psi_B\rangle|\nn\\
&\leq\; \sqrt{2}.
\end{align}

\end{document}